\documentclass[aps,pre,onecolumn,superscriptaddress]{revtex4}

\usepackage[dvips]{graphicx}
\usepackage[dvips]{color}
\usepackage{hyperref,breakurl,amsmath,amssymb,pst-node,eepic}
\usepackage{graphicx}% Include figure files
\usepackage{dcolumn}% Align table columns on decimal point
\usepackage{bm}% bold math
\usepackage{hyperref}
\usepackage{float}
\usepackage{xcolor}

\begin{document}

\preprint{APS/123-QED}

\title{Who should fight the spread of fake news?}% Force line breaks with \\

\author{Diana Riazi}
 %\altaffiliation{Department of Computer Science, University College London.}%Lines break automatically or can be forced with \\
%\author{}%
\affiliation{Department of Computer Science, University College London, 66-72 Gower Street, WC1E 6EA London, United Kingdom}

\author{Giacomo Livan}
\affiliation{Dipartimento di Fisica, Universit\`a degli Studi di Pavia, Via A. Bassi 6, 27100 Pavia, Italy}
\affiliation{Department of Computer Science, University College London, 66-72 Gower Street, WC1E 6EA London, United Kingdom}


%\affiliation{
%\affiliation{
%\affiliation{%


\date{\today}% It is always \today, today,

\begin{abstract}
This study investigates who should bear the responsibility of combating the spread of misinformation in social networks. Should that be the online platforms or their users? Should that be done by debunking the `fake news' already in circulation or by investing in preemptive efforts to prevent their diffusion altogether? We seek to answer such questions in a stylized  opinion dynamics framework, where agents in a network aggregate the information they receive from peers and/or from influential external sources, with the aim of learning a ground truth among a set of competing hypotheses. In most cases, we find centralized sources to be more effective at combating misinformation than distributed ones, suggesting that online platforms should play an active role in the fight against fake news. In line with literature on the `backfire effect', we find that debunking in certain circumstances can be a counterproductive strategy, whereas some targeted strategies (akin to `deplatforming') and/or preemptive campaigns turn out to be quite effective. Despite its simplicity, our model provides useful guidelines that could inform the ongoing debate on online disinformation and the best ways to limit its damaging effects.

\end{abstract}

\maketitle

%\tableofcontents

\section{\label{sec:level1}Introduction}

Having seemingly arrived to a post-truth era~\cite{lewandowsky2017beyond} guided by propaganda and ideological polarization~\cite{spohr2017fake}, we have witnessed the apparent commonality with which misinformation propagates in social networks and thus current and future prospective risks of infodemics~\cite{gallotti2020assessing}. Specifically, it has now been well established that `fake news' spreads faster as compared to pieces of true news~\cite{hornik2015information}, in line with the fact that many individuals possess an inclination towards sensationalism~\cite{prior2003any}. Over recent years in particular, fake news has demonstrated its potential to lead to social unrest and to threaten and destabilize democracies~\cite{zhang2020overview}.

What are or have been the attitudes and efforts practiced by social media platforms thus far? X (formerly Twitter) for instance had labelled content, prompting users when engaging or interacting with a tweet which had been deemed `misleading', thereby contextualizing such tweets.  Additionally, the social media platform had created `Twitter moments', though this has now also been removed, where users had the opportunity to `learn from other users on Twitter and trusted sources', where essentially certain tweets had been deemed noteworthy in some way, favored by the algorithm in place or by the curation team~\cite{duguay2018social}. At the moment, X has instead now launched `community notes,' also with the expressed aims of adding context to posts and keeping users informed, with the platform indicating that this feature is not written, rated or moderated, unless they go against platform regulations \footnote{https://communitynotes.x.com/guide/en/about/introduction}. After being criticized for not doing enough to tackle misinformation spread occurring on its platform, shortly following the 2016 US presidential election, Meta initiated  agreements with external fact-checking firms in an effort to manage the amount of misinformation \footnote{https://www.facebook.com/formedia/blog/third-party-fact-checking-how-it-works} in addition to affirming that content which is deemed misleading is removed \footnote{https://transparency.meta.com/en-gb/policies/community-standards/misinformation/}. 

As it is known, practiced methods of combating misinformation spread in online social networks include debunking \cite{helfers2023differential} --- actively promoting a ground truth in the face of circulating misleading information --- and prebunking \footnote{https://www.washingtonpost.com/business/2021/11/01/twitter-climate-disinformation/}, i.e. the attempt to preemptively establish a ground truth. Much of the literature regarding whether misinformation correction occurring via debunking shows nuanced results regarding its effectiveness. It has been established individuals often resist against information correction, giving rise to the so called `backfire' effect~\cite{zollo2017debunking}. At the same time, debunking has been shown to be effective, though to not completely erase the effects of misinformation, in a scenario where there exists a repetition of retraction~\cite{ecker2011correcting}, or when it comes from a trusted source~\cite{vinck2019institutional, xiao2021dangers}. Moreover, an individual's ideological predisposition~\cite{zollo2017debunking,jang2019debunking} plays a critical role when they are presented with correcting information, ultimately resulting in either their acceptance or resistance. This suggests that the timing with which correcting information is presented may be crucial, suggesting in turn that prebunking may provide valuable solutions to countering the effects of misinformation.

In addition to debunking and prebunking, deplatforming, i.e. the removal or suspension of a user's account if they have been identified to have violated a social media platform’s community terms/regulations \footnote{https://transparency.meta.com/en-gb/policies/community-standards/hate-speech/}, is a practiced procedure to lower misinformation levels~\cite{jhaver2021evaluating, klinenberg2024does}.
Most recently, it has been reported that the  riots following the mass stabbing in Southport (UK) had been fueled by far-right communities on very large social media platforms, such that the UK had reportedly considered forcing tech firms to remove `legal but harmful' content after riots \footnote{https://www.ft.com/content/6b886570-1b55-4647-8b6a-c4ad4c5b6925}. At the same time, it is necessary to note evidence of cross-platform migration and emergence of toxic and extreme communities being consequences and associated risks of this practice~\cite{ali2021understanding}. 

Beyond the above more conventional counter-measures, so called inoculation against misinformation, i.e. the repeated exposure to `small doses' of misinformation, has been shown within the psychology literature to build resistance and `immunity' against stronger strains of `fake news'~\cite{lewandowsky2021countering,cook2017neutralizing}.

When evaluating the effectiveness counter-strategies against misinformation, it is necessary to distinguish among the sources from which such misinformation may originate (e.g., a charismatic influential figure, or a distributed group of social media users) as well as the potential sources of correction. Here we aim to investigate the ethically-underlying and pragmatic question of who should bear the responsibility of combating misinformation, that is, who or what should be societies’ relied upon source of  mis/dis-information correction. Should that be centralized sources such as governments or social media platforms themselves, or should that instead be decentralized populations of individuals and/or platform users?

To gauge this question, we invoke a social learning procedure known as Distributed Hypothesis Testing (DHT) as introduced in~\cite{lalitha2018social} and further developed  in~\cite{riazi2024public,riazi2024mitigating}, which allows us to easily quantify quantities that are key to assessing a population's resilience against misinformation. The most important one of them is \textit{truthfulness}, i.e., the overall alignment between a population's average opinions and a ground truth. On top of that, our framework allows us to keep track of both the publicly stated and privately held beliefs of agents in a population, potentially leading to permanent misalingments between the two, a feature which is reminiscent of the psychological phenomenon usually referred to as \textit{cognitive dissonance}~\cite{cooper2019cognitive} (CD). 

In the following, through numerical simulations, we seek to compare the effects of stylized centralized and decentralized sources of disinformation as well as sources of correcting interventions (i.e., debunking, prebunking, deplatforming, and inoculation) on both truthfulness and CD, while gauging the potential asymmetry between such misinforming and correcting forces. We run our simulations on networked populations, as network topology (e.g., homogenoeus vs heterogeneous) is known to have a significant impact on the diffusion of (mis)information~\cite{stern2021impact}.

\section{\label{sec:level3}Social Learning and Distributed Hypothesis Testing}
%%%%%%
The objective in DHT models is for agents to learn a ground truth, given an array of competing hypotheses or narratives, through the compilation of privately-held signals over time. Such signals are obtained as random draws from a distribution based on the ground truth, though agents do not possess awareness of this and the learning procedure holds the aim of deciphering from which distribution they are receiving signals among an array of different distributions.  Distributions may be viewed to capture  the uncertainty which surrounds such hypotheses or perspectives. In the following, a concise summarization of the DHT process as was first introduced by ~\cite{lalitha2018social} as well as recall the notions of truthfulness and cognitive dissonance, as introduced and utilized in ~\cite{riazi2024public,riazi2024mitigating}.  
%%%%%%%
We consider $N$ agents connected by a network $W$ of interactions (such that $W_{ij} > 0$ when agents $i$ and $j$ are connected and $W_{ij} = 0$ otherwise). The weight $W_{ij}$ represents the amount of influence that agent $i$ accepts from agent $j$. Because of this, the matrix $W$ is assumed to be row-stochastic, i.e. $\sum_{j=1} ^{N} W_{ij}=1$. Let us also consider a set of $N \times M$ multivariate distributions $f_i(X; \theta_{ik})$ ($i = 1, \ldots, N$, $k = 1, \ldots, M$), where $\theta_{ik}$ denotes the set of parameters that define the $k$-th distribution associated with agent $i$. To simplify notation, in the following we will drop the subscript $i$, as it will be clear from context to which agent the set of parameters $\theta_{ik}$ refers to.

The $M$-th distribution is the one corresponding to the ground truth for each agent. Time is discrete and denoted as $t=1,...,T$. At time $t = 0$ each agent is initialized with a random vector of private beliefs $\boldsymbol{q}_i^{(0)} = \left (q_i^{(0)}(\theta_1), \ldots, q_i^{(0)}(\theta_M) \right )$, such that $q_i^{(0)}(\theta_k) \geq 0, \ \forall k$ and $\sum_{k=1}^M q_i^{(0)}(\theta_k) = 1$. At each time step, we perform the following steps: 

\begin{itemize}
    \item Each agent $i$ ($i = 1, \ldots, N$) receives a signal 
    $X_i^{(t)}$ as a random draw from the distribution corresponding to the ground truth, i.e., $X_i^{(t)}\sim f_i (\cdot ; \theta_M )$.
    \item Each agent $i$ performs a local Bayesian update on their current vector of private beliefs $\boldsymbol{q}_i^{(t)}$ to form a public belief vector $\boldsymbol{b}_i^{(t)}$ with components 
    \begin{equation} \label{eq:pub_beliefs}
        b_{i}^{(t)}(\theta_k) = \frac{f_{i}(X^{(t)}_{i}; \theta_k) \ q_i^{(t-1)}(\theta_k)}{\sum_{\ell = 1}^M f_i(X^{(t)}_{i};\theta_\ell) \ q^{(t-1)}_{i}(\theta_\ell)}
    \end{equation}            
    \item Each agent $i$ shares their public belief vector $\boldsymbol{b}_i^{(t)}$ with all their neighbors in the network, and similarly receives public belief vectors from each of them.
    \item Each agent $i$ updates their private belief vector $\boldsymbol{q}_i^{(t)}$ by averaging the log-beliefs they received from neighbors, i.e.,
    \begin{equation} \label{eq:pvt_beliefs}
q^{(t)}_{i}(\theta_k)=\frac{\exp \left (\sum_{j=1}^N W_{ij} \log b^{(t)}_{j}(\theta_k) \right)}{\sum_{\ell = 1}^M \exp \left (\sum_{j=1}^N W_{ij}\log b^{(t)}_{j}(\theta_\ell) \right )} \ .
    \end{equation}
\end{itemize}
Note that both Eqs.~\eqref{eq:pub_beliefs} and~\eqref{eq:pvt_beliefs} ensure that the public and private belief vectors of each agent remain correctly normalized as probability vectors at each time step.

As shown in~\cite{lalitha2018social}, the ground truth is collectively learnt by all agents exponentially fast under global distinguishability, which characterizes the concept that at least one agent in the network is able to differentiate between any pair of competing hypotheses, i.e., for all $k \neq \ell$, there exists at least one agent $i$ such that $D_\mathrm{KL}(f_i(\cdot,\theta_k) || f_i(\cdot,\theta_\ell)) > 0$, where $D_\mathrm{KL}(\cdot || \cdot)$ denotes the Kullback-Leibler divergence.

To lend some intuitiveness to the model, let us consider the following scenario. Let us suppose that the agents in our model are discussing the effectiveness of vaccines and the myth that they cause cases of autism \cite{davidson2017vaccination}. For instance, we may consider the set of mutually exclusive hypotheses discussed by the agents to be as follows. $\theta_1$ may correspond to the perspective that `vaccines are not effective in reducing transmission and is linked to autism'; $\theta_2$ to `vaccines are effective in reducing transmission but may be a cause of cases of autism'; $\theta_3$ to `vaccines are effective in reducing transmission and are not linked to autism' (ground truth).

Overall, the $\theta$s may be viewed as competing perspectives which may be affirmed by agents, and distributions $f_i$s represent the uncertainty surrounding such perspectives. The signals the agents draw from the distribution $f(\theta_M)$  represent instances of information oriented with the ground truth. In our vaccine analogy, one such signal could be represented by an agent's observation, e.g., an individual contracting an illness and recovering. In this respect, the agents' task is therefore trying to figure which perspective this piece of information supports. In order to do that, agents communicate with other agents through their social network in order to aggregate information and ultimately come to the ground truth.

The aforementioned notion of distinguishability captures the fact that the agents may confuse different hypotheses given to, e.g., the convoluted nature of the topic being discussed or relevant idiosyncrasies that the model is not able to take into account (e.g., political orientation, background, etc.). In our analogy, for instance, agents opposed to vaccine mandates/regulations may conflate the hypotheses that a vaccine is ineffective in reducing transmission or that it is effective in reducing transmission but is linked to autism.
%%%%%%%%%%%%%%%%%

In the next section, we recall our notions of truthfulness and CD as introduced and utilized in \cite{riazi2024public,riazi2024mitigating}. 

\section{\label{sec:level4}Measures of Truthfulness and Cognitive Dissonance}

To characterize the overall network's ability to learn the ground truth, we make use of the following metric introduced in~\cite{riazi2024public}:
 
\begin{equation} \label{eq:truthfulnes}
    \tau(t) =\frac{1}{N}\sum_{i=1}^N q_i^{(t)}(\theta_M) \ ,
\end{equation}
which quantifies the average private belief placed by the agents on the ground truth (recall that by convention the assumption of the $M$-th hypothesis to be the true one). The quantity in Eq.~\eqref{eq:truthfulnes} is referred as \emph{truthfulness}. Truthfulness can also be expressed as the difference between one and the average private belief collectively placed on the $M-1$ wrong hypotheses due to the normalization of private vectors:
\begin{equation} \label{eq:CD}
    \tau (t)= 1 - \frac{1}{N}\sum_{i=1}^N \sum_{\ell=1}^{M-1}
    q_i^{(t)}(\theta_\ell) \ .
\end{equation}

The concept of  cognitive dissonance (CD) in Psychology refers to the mental toll experienced by an individual when faced with contradictory information~\cite{cooper2019cognitive}. In the context of our model, we equate CD to the difference (in absolute value) between what an agent privately believes and what they publicly express. Mathematically, the CD experienced by an agent $i$ at time $t$ regarding some hypothesis $\theta_k$ is
\begin{equation}
C^{(t)}_i(\theta_{k})= \left |q^{(t)}_i(\theta_k)-b^{(t)}_i(\theta_k) \right | \ .
\end{equation}

\section{Centralized versus Decentralized}

Our intention is to understand who or what should bear the responsibility of countering the spread of misinformation in social networks. Is it better dealt with from a centralized source, such as governments, influential figures/leaders, social media platforms themselves? Or is it more pragmatically managed from decentralized, distributed sources i.e. journalists, scientists, experts, truth-seeking individuals?

We consider simulating both centralized and decentralized sources of disinformation as well as sources of information correction within our DHT framework. While of course ethics in part govern the question at hand, though such arguments are not made here, this paper intends to evaluate this by gauging the effects dynamically within the DHT framework. This classification of sources of information into centralized and decentralized can assist in illuminating potentially generalizable outlets towards the management and moderation of misinformation spread in social networks. 

To imitate centralized sources of information (e.g. governments, social media platforms, influencers), we introduce, what we refer to as `mega-nodes.' A mega-node bolsters their probability belief vector in each time-step. At each time step, each agent $i$ ($i=1 \ldots N$) listens to a given mega-node with probability proportional to the dot product between their private belief vector and the selected mega-node's belief vector, i.e., with probability $\boldsymbol{q}_{i}^{t} \cdot \boldsymbol{b}_{mg}^{t}$, where $\boldsymbol{b}_{mg}=(b_1,b_2,b_3,\ldots)$ with $b_i \approx 1$ and $b_j \ll 1$ for $j \neq i$. Such a probabilistic rule captures in a stylized manner well known cognitive biases, such as the tendency to seek information from sources that are aligned with one's preexisting beliefs~\cite{sikder2020minimalistic}, the resistance to information correction~\cite{ecker2022psychological}, or the consequences of ideological polarization~\cite{vegetti2019political, spohr2017fake}.

We consider a debunker mega-node pushing the ground-truth, i.e., a node whose belief vector at all times is such that $b_M^{(t)} \approx 1$ and $b_j^{(t)} \ll 1$ for $j = 1 \ldots, M-1$, and a conspirator mega-node with $b_1^{(t)} \approx 1$ and $b_j^{(t)} \ll 1$ for $j = 2 \ldots, M$. The latter may be seen, for instance, as an influential political figure promoting sensationalized unsubstantiated claims. The debunking counterpart may instead be likened to centralized sources such as governments or influential figures grounded in scientific evidence attempting to demystify a demonstrable fact, as well as social media platforms themselves implementing `fact-checking' efforts.
%literature on opinion leaders

For the decentralized case, we recall the `two-tribe' model originally introduced and discussed in~\cite{riazi2024public}. More specifically, in addition to `regular' agents who update their beliefs according to DHT as described above, there are two distinct sub-populations of distributed agents, namely \textit{conspirators} and \textit{debunkers}, where the former can be interpreted as proxy for disinformation flow, in that such agents statically promote a non-truth, and the latter to be agents who promote the ground-truth. We denote the fraction of such distributed conspirator and debunker agents with $\beta_c$ and $\beta_d$ respectively. More specifically, it is imposed that conspirator agents maintain a public belief vector $\boldsymbol{b}^{(t)} = (b_1,b_2,b_3,\ldots)$ with $b_1 \approx 1$ and $b_j \ll 1$ for $j = 2,\ldots, M$, that is, conspirator agents push one of the wrong hypotheses (which,  once again, without loss of generality, we assume to be the first one) at each time step. Similarly, decentralized debunker agents promote the ground-truth such that, as in the centralized case, $b_M^{(t)} \approx 1$ and $b_j^{(t)} \ll 1$ where $j=1 \ldots M-1$. This scenario may be analogized to individual, distributed agents that misinform (e.g. malicious bots) or correct misleading information circulating social networks (e.g. experts, journalists, members of the scientific community).

It should be noted that the line between centralized and decentralized may not be so clean-cut in actuality, or that centralized and decentralized forces may evidently co-exist. In fact, one form of information origin may have the potential to inform the other. However, we do not consider the co-existence of centralized and decentralized together. Rather, by classification of sources of information, the dominant interest of this paper is to come towards clarifying the magnitude of effects associated with centralized and decentralized sources of information. 

In the next section, we seek to compare centralized and decentralized sources of information as described and if there are any notable differences on their effect on our measures of truthfulness and CD. This is in an effort to ultimately gauge, given a centralized or decentralized source of mis/dis-information, what entity is best at regulating the spread of online mis/dis-information spread, if at all,  i.e. who should fight the spread of `fake news.'

\section{Results}
We numerically simulate a series of networks under both the centralized and decentralized iterations described above. In an effort to elucidate what type of source of information is better at moderating the spread of misinformation, we break down our approach into the following sub-questions: 
\begin{itemize}
    \item \textbf{Q1}: \textit{Given a centralized source of information (mega-nodes), what are the dynamics of truthfulness and CD, considering the presence and absence of conspiring and debunking mega-nodes?}
   \item \textbf{Q2}: \textit{Is centralized or decentralized debunking more effective in achieving truthfulness?}
   \item  \textbf{Q3}: \textit{Is prebunking more or less effective than debunking in achieving higher truthfulness?}
   \item \textbf{Q4}: \textit{Does there exist an asymmetry between conspiring and debunking centralized sources (mega-nodes) in the strengths with which their respective agendas are pushed?}
   \item \textbf{Q5}: \textit{Are there differences in truthfulness between deplatforming conspiring sources of information and not deplatforming at all?}
   %Is deplatforming conspiring sources of information effective in harnessing a collective belief in the ground-truth as compared to not? 
   \item \textbf{Q6}: \textit{Is psychological inoculation effective in achieving higher levels of truthfulness as well as reducing CD in comparison to no such treatment?}
  
\end{itemize}

We begin our investigation with witnessing the effects within Erdos-Rényi (ER) networks, which in the limit of large network size are characterized by a Poissonian (i.e., homogeneous) degree distribution. While ER networks do not capture heterogeneity of the degree distributions of real-world social networks, they provide adequate and sufficient means in initially comparing our described centralized and decentralized methods of disinformation spread. We ultimately make considerations in heterogeneous Barabasi-Albert networks~\cite{barabasi1999emergence} (BA), characterized by a long-tailed degree distribution, as later addressed in this section. 

\begin{figure}[h!]
    \centering
       \includegraphics[width=7cm]{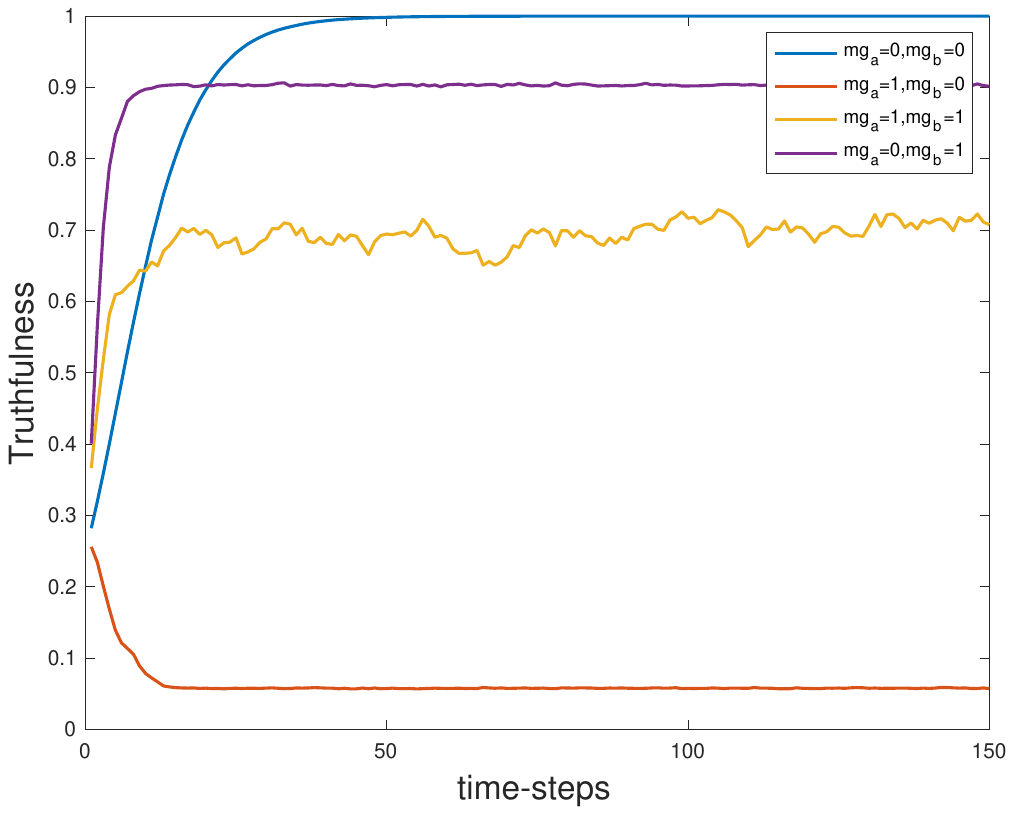}
      \includegraphics[width=7cm]{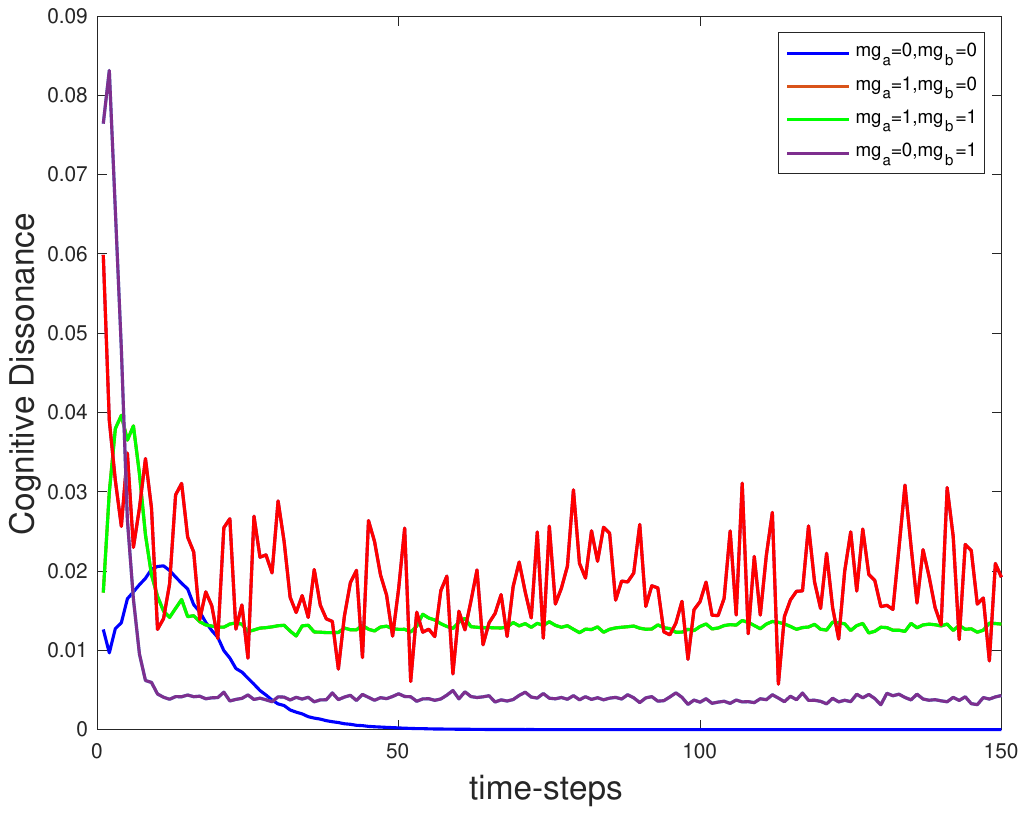}
    \caption{Temporal evolution (left) of Truthfulness, (right) of CD, under `mega-nodes.' Results are obtained by averaging over $100$ independent simulations run on sparse ER networks with $N=100$ and $M=4$. Legend: $mg_a$ and $mg_b$ refers to conspiring and debunking mega-node respectively where $0$ denotes absence and $1$ presence }
    \label{fig:1}
\end{figure}

Firstly, in reference to \textbf{Q1} we seek to compare, given a centralized origin of information, conspiring and debunking regimes, and gauge their impact on the DHT dynamics.
In Fig.~\ref{fig:1}, we observe the temporal evolution of truthfulness and CD under different regimes of mega-nodes gauged against `unperturbed' DHT dynamics~\cite{lalitha2018social}. Intuitively, some degree of truthfulness is recovered through the addition of a `debunking' mega-node to counter the damage inflicted by a conspirator mega-node. We make note of the legend utilized, where $mg_a$ and $mg_b$ refer to a conspiring and debunking mega-node respectively where $0$ denotes absence and $1$ presence. In the right panel of Fig.~\ref{fig:1}, in the standard DHT case ($mg_a=0, mg_b=0$) we observe the subsiding of the agents' collective CD after a sufficient number of time-steps. However, there are noticeably more volatile levels of CD in the presence of the `conspiring' mega-node alone ($mg_a=1, mg_b=0$), as well as alleviating effects delivered by the `debunking' mega-node ($mg_b=0$). Notably, `the back-fire effect'~\cite{zollo2017debunking} is captured by the fact that the scenario in which only the `debunker' mega-node is active ($mg_a=0, mg_b=1$) is strictly worse than the one in which no mega-node is active, both in terms of  truthfulness and CD.

Coming to \textbf{Q2}, we move towards distinguishing any differences in effects between centralized and decentralized debunking forces. Ultimately, we desire to determine here whether centralized or decentralized debunking is more effective in increased truthfulness. Recall, as described in the previous section, that a centralized source within our framework refers to agents being `predisposed' to listening and thus adopting the belief vector, publicly and privately, of a mega-node, whereas decentralized refers to a sub-population of dispersed agents throughout a network communicating a belief vector in each time-step, pushing a particular hypothesis. 
 
\begin{figure}[h!]
    \centering
      \centerline{
       \includegraphics[width=7cm]{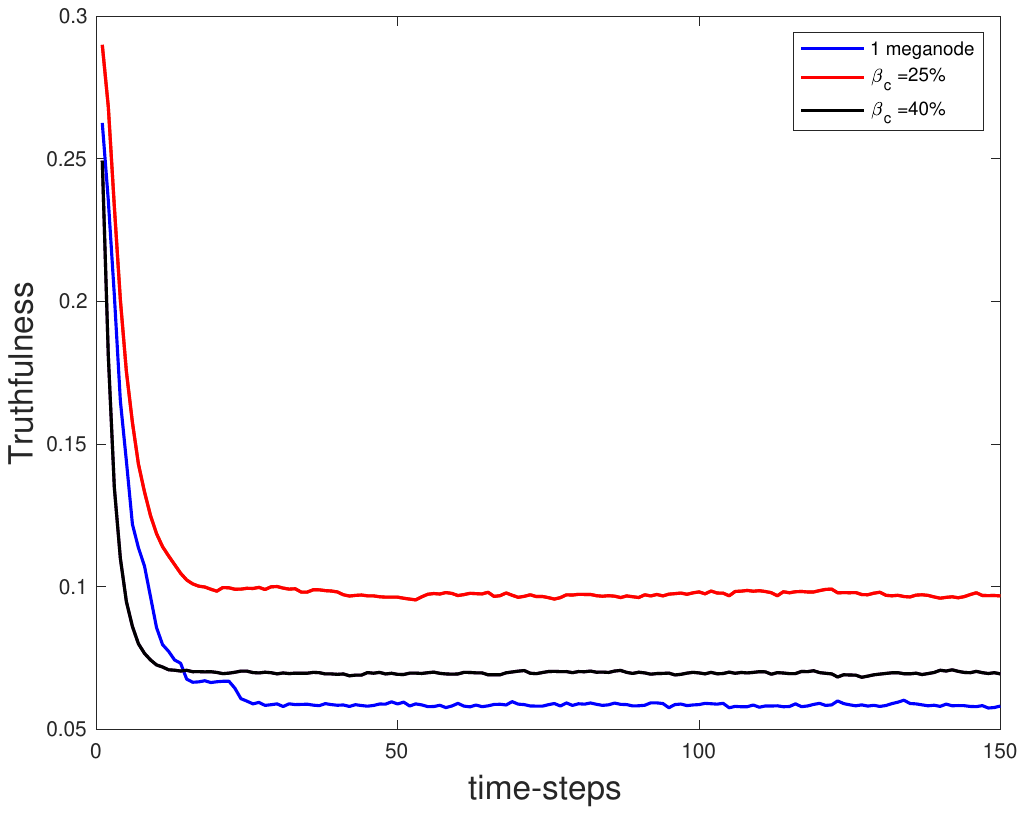}
       \includegraphics[width=7cm]{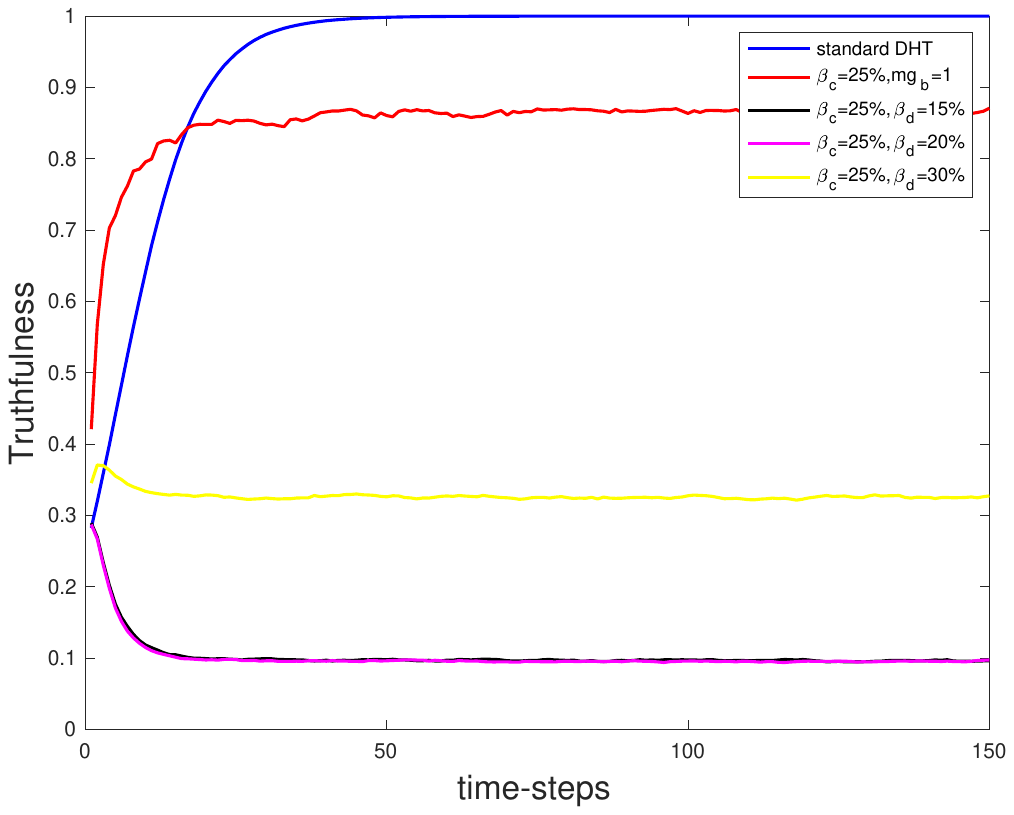}}
    \caption{Left: Temporal evolution of truthfulness when disinformation originates from a centralized source (mega-node) or a sub-population of distributed conspirator agents amounting to a fraction $\beta_c$ of the whole population. Right: Temporal evolution of truthfulness in the presence of debunking from both centralized (one mega-node, red line) and decentralized (a fraction $\beta_d$ of the population) sources against disinformation originated from a decentralized source (a fraction $\beta_c$ of the population). Results are obtained by averaging $100$ independent simulations run on sparse ER networks with $N=100$ and $M=4$.}
    \label{fig:2}
\end{figure}

\begin{figure}[h!]
    \centering
      \centerline{
      \includegraphics[width=7cm]{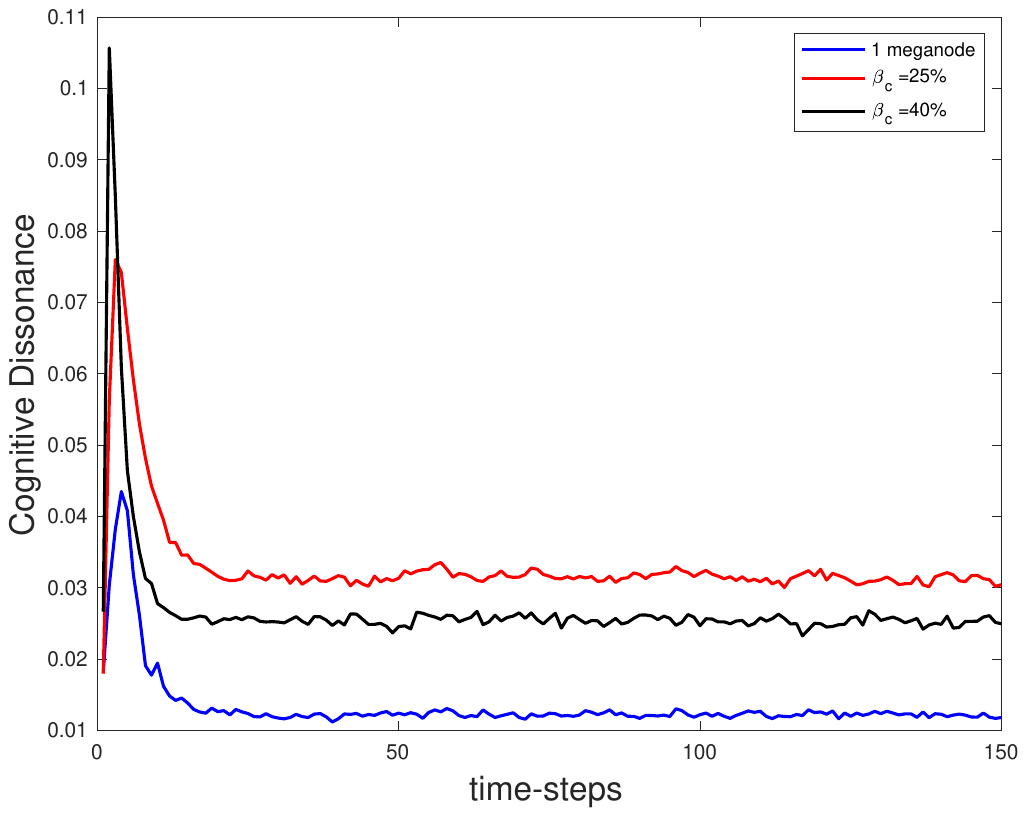}
    \includegraphics[width=7cm]{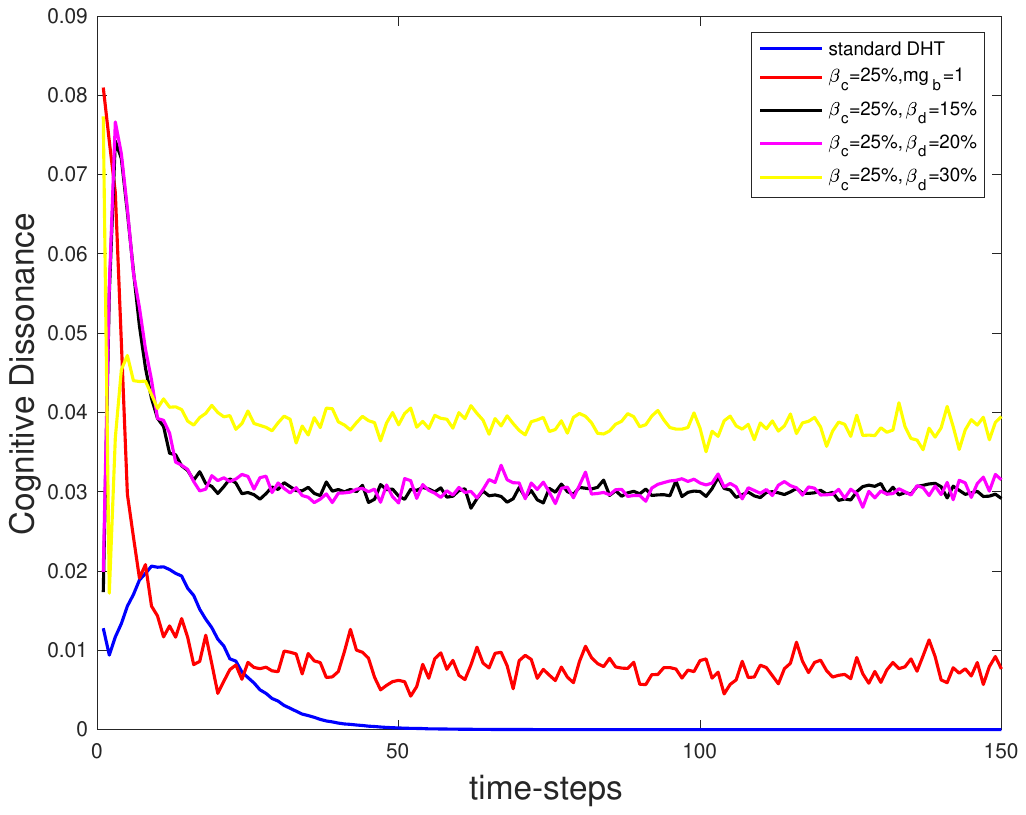}}
    \caption{Left: Temporal evolution of CD when disinformation originates from a centralized source (mega-node) or a sub-population of distributed conspirator agents amounting to a fraction $\beta_c$ of the whole population. Right: Temporal evolution of CD in the presence of debunking from both centralized (one mega-node, red line) and decentralized (a fraction $\beta_d$ of the population) sources against disinformation originated from a decentralized source (a fraction $\beta_c$ of the population). Results are obtained by averaging $100$ independent simulations run on sparse ER networks with $N=100$ and $M=4$. }
    \label{fig:4}
\end{figure}

In Figs.~\ref{fig:2} and~\ref{fig:4}, we compare centralized and decentralized sources of disinformation and debunking. More specifically, we observe in the left panel of Fig.~\ref{fig:2} that a centralized source of disinformation is more effective in delivering lower levels of truthfulness compared to sizeable concentrations of decentralized conspirator agents. In the right panel of Fig.~\ref{fig:2} we see that, given a decentralized source of disinformation, a centralized debunking mega-node is better at building resilience against a conspirator mega-node than distributed debunker agents. In Figure~\ref{fig:4}, we notice that such dynamics are mirrored in the CD counterpart of centralized and decentralized comparisons. 
 
 We now seek to understand if \textit{prebunking}, i.e. preemptively promoting the ground truth before any entry of misleading information, has any impactful effects on boosting the agents' collective belief in the truth, as posed in \textbf{Q3}. To simulate this, we begin with a `prebunking' sub-period of $T_p=10$, where the ground-truth is promoted by a centralized source, that is a `prebunking' mega-node, before the entry of any conspiring source which misinforms. Within this sub-period of $T_p$, during any time step the agents are randomly selected to listen to the prebunking mega-node with probability equal to the dot product between their private belief vector and the prebunking mega-node's public belief vector. In other words, agents are predisposed to listening to the prebunking mega-node during each time-step of the prebunking sub-period. 
 For the decentralized counter-part of prebunking, similar to that of distributed debunkers, a fraction of distributed prebunkers, denoted by $\beta_p$ where within the prebunking sub-period $T_p$, such agents preemptively promote the ground truth, where $b_M^{(t)} \approx 1$ and $b_j^{(t)} \ll 1$ where $j=1 \ldots M-1$. From $T_p$ onwards, a conspiring sources or source may enter and misinform through the promotion a non-truth, as outlined in previous cases. We note that the prebunking source(s) is/are evidently only active within the prebunking sub-period.

\begin{figure}[h!]
    \centering
      \centerline{
      \includegraphics[width=7cm]{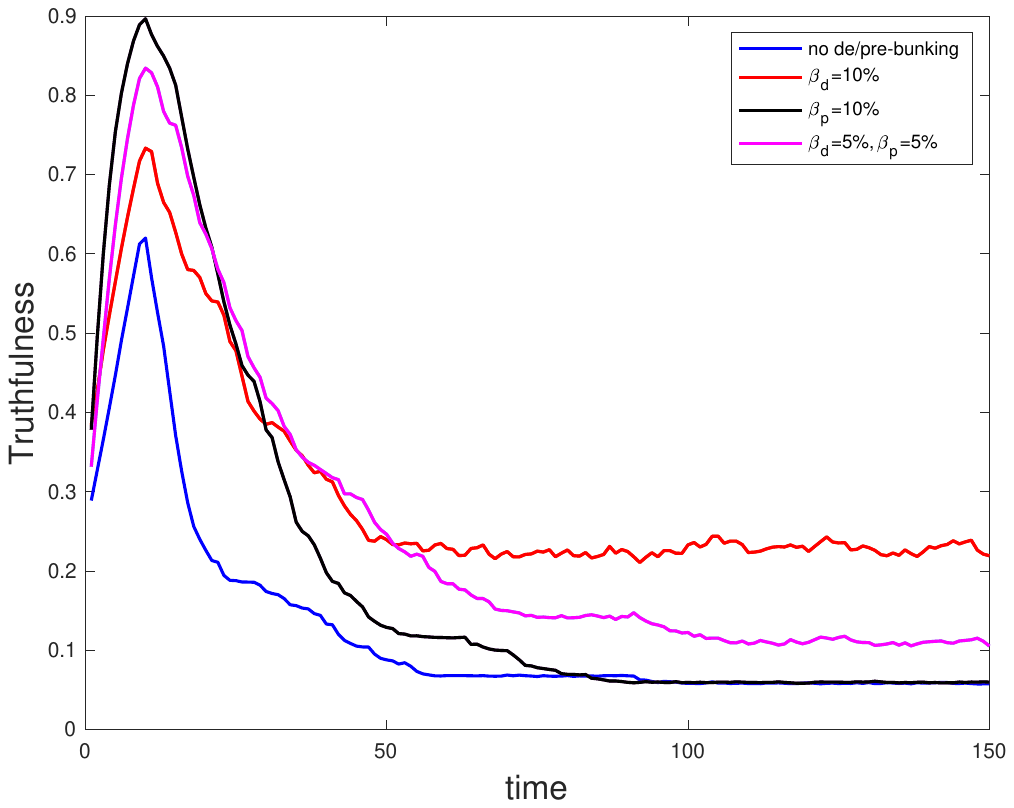}
      \includegraphics[width=7cm]{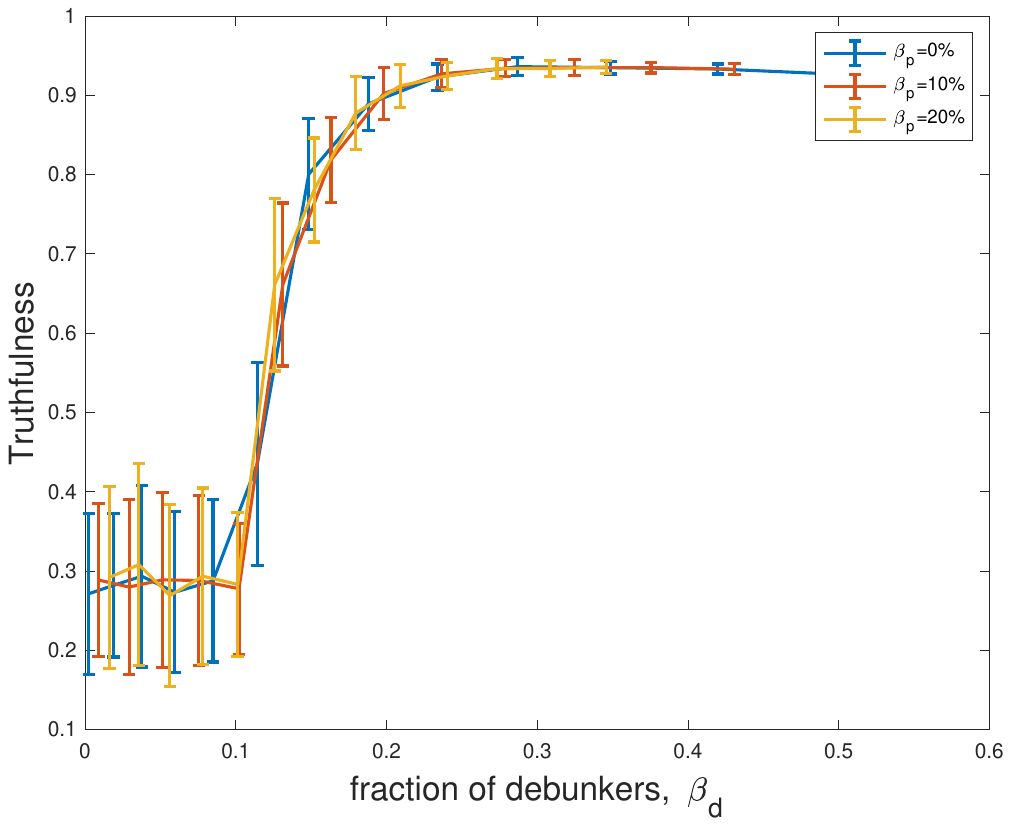}}
    \caption{Left: Temporal evolution of truthfulness in scenarios with decentralized debunking and prebunking, carried out by a fraction $\beta_d$ and $\beta_p$ of the population, respectively. Right: steady-state truthfulness obtained in simulations with different concentrations $\beta_p$ of decentralized prebunker agents as a function of the concentration $\beta_d$ of debunker agents where errorbars capture the differences across simulations. Results are obtained by averaging $100$ independent simulations run on sparse ER networks with $N=100$ and $M=4$. }
       \vspace*{8pt}
    \label{fig:5}
\end{figure}

\begin{figure}[h!]
    \centering
      \centerline{
      \includegraphics[width=7cm]{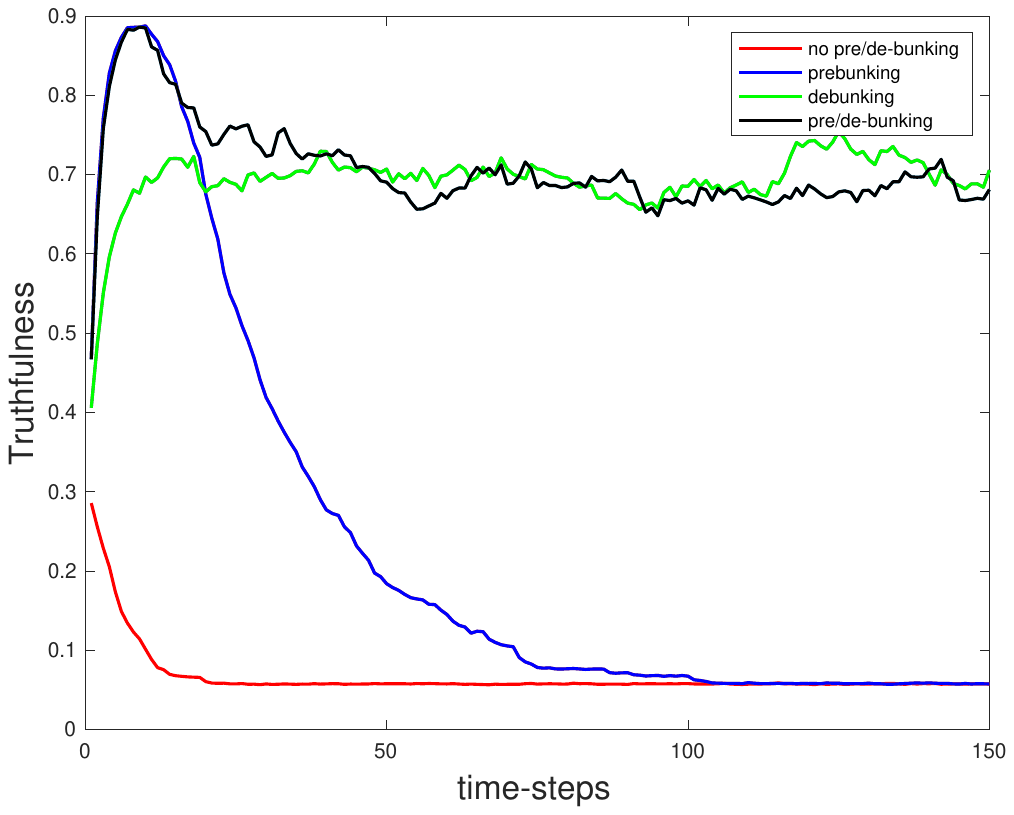}}
    \caption{Temporal evolution of truthfulness in scenarios with/without centralized debunking and/or prebunking. Results are obtained by averaging $100$ independent simulations run on sparse ER networks with $N=100$ and $M=4$. }
     \label{fig:6}
\end{figure}

In Figs~\ref{fig:5} and~\ref{fig:6}, it can be seen that the effects of prebunking die out, such that ultimately the arrived upon steady-state is effectively comparable with no interventions which promote the ground-truth (i.e., no prebunking or debunking). These observations are in line with the literature on prebunking, which, as mentioned, suggests mixed results \cite{tay2022comparison, jolley2017prevention}. The idea of prebunking is to arm agents proactively with the ground truth, where deviations from it may be anticipated. However, according to our results, there is a possibility for agents to `unlearn' the ground truth upon the entrance of a conspiring force.

While we began our investigation within ER networks, we provide equivalent simulations in Barabási-Albert (BA) networks with Figs.~\ref{fig:9}-\ref{fig:11} (see Appendix), as BA networks represent a standard for a heterogeneous, heavy-tailed, degree distributions \cite{barabasi2003scale} where notably differences between centralized and decentralized forms of debunking are less pronounced compared to that of ER simulations.

Thus far, we have assumed a constant `strength' or probability vector with which each mega-node pushes their respective agenda. However, we may consider varying this, capturing the manner in which a topic (at times regarding a demonstrable fact~\cite{mccright2011politicization,wang2019systematic}, such as the reality of climate change or the effectiveness of vaccines) is covered by a variety of sources, such as say news outlets, which are often associated with a certain political leaning or possessing an aim of delivering a `balanced story'~\cite{lewandowsky2012misinformation}. This could be due to a range of motivations, such as the aim of appearing more neutral in an effort to maximize engagement from an ideologically diverse set of users. With these points in mind, we now tackle \textbf{Q4}. 

We denote $\rho \in (0,1)$ to be the probability with which the debunking mega-node pushes the ground-truth, with the remainder of beliefs regarding $\theta_k, k=1 \ldots M-1$ determined via random draws from a uniform distribution rescaled to ensure normalization  i.e. the debunking mega-node's public belief vector is such that $\boldsymbol{b}_{i}=(b_1,b_2,b_3 \ldots)$, where $b_M=\rho$ such that $\sum_k b_{i} =1$.  Similarly, we denote $\rho_m\in (0,1)$ to be the probability with which the conspiring mega-node pushes their agenda, i.e., a non-truth (for the sake of consistency in simulations, we have assumed this to be $\theta_1$), with the remainder of beliefs regarding $\theta_k, k=2 \ldots M$ determined via random draws from a uniform distribution rescaled to ensure normalization, i.e. the conspiring mega-node's public belief vector is such that $\boldsymbol{b}_{i}=(b_1,b_2,b_3 \ldots )$, where $b_1=\rho_m$ such that $\sum_k b_{i} =1$.

\begin{figure}[h!]
    \centering
      \centerline{
      \includegraphics[width=7cm]{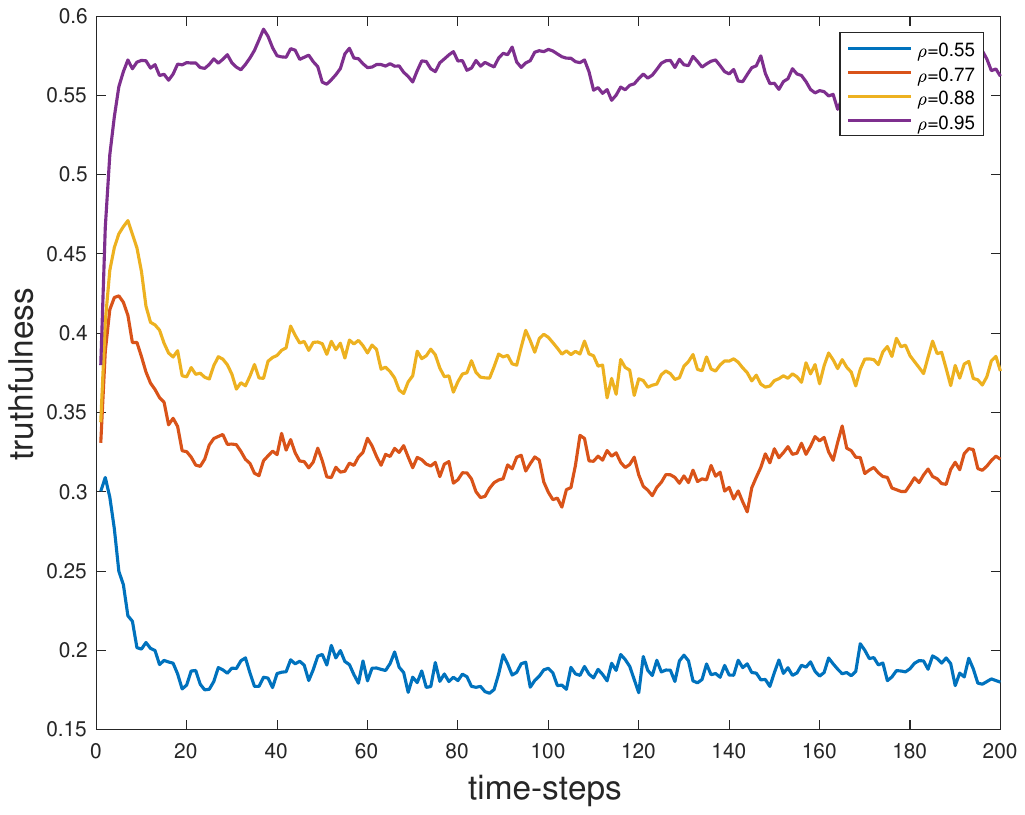}
      \includegraphics[width=7cm]{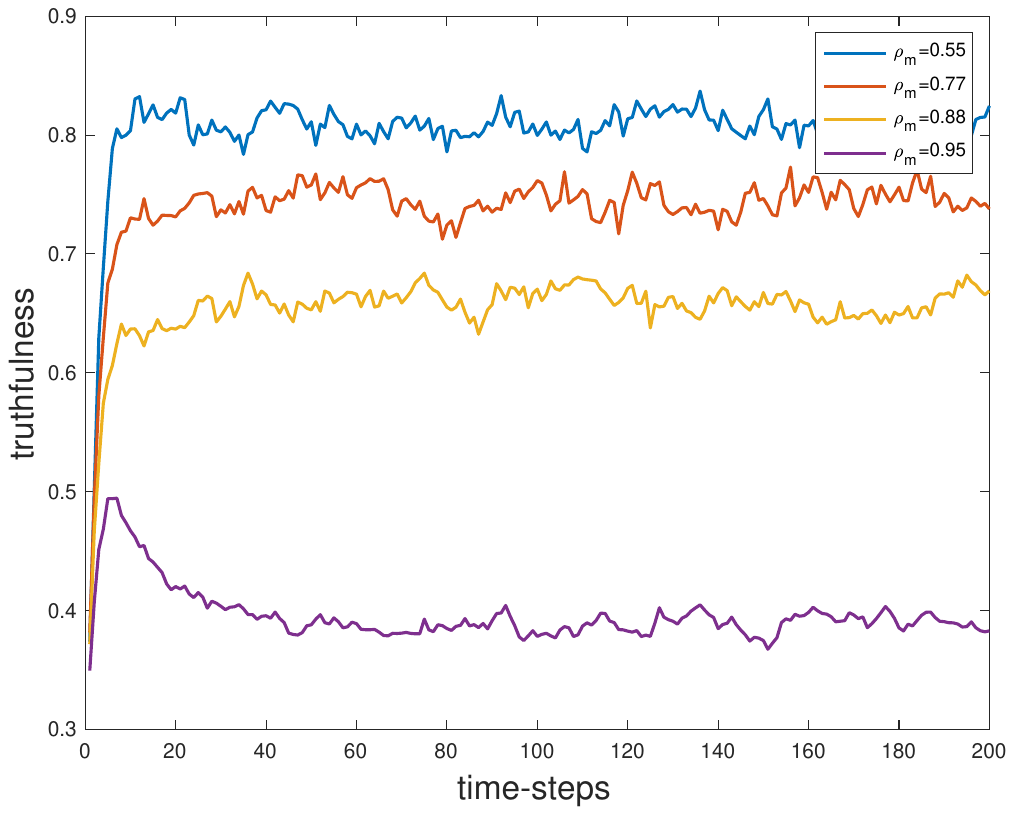}}
    \caption{Temporal evolution of truthfulness for (left panel) an array of values of $\rho$ --- the probability with which the debunking mega-node pushes the ground-truth --- with a constant $\rho_m\approx 1$, and (right panel) an array of values of $\rho_m$ --- the probability with which the conspiring mega-node pushes their agenda --- with a constant $\rho \approx 1$. Results are obtained by averaging $100$ independent simulations run on sparse ER networks with $N=100$ and $M=4$. }
    \label{fig:12}
\end{figure}

In the left panel of Fig.~\ref{fig:12}, we consider different values of $\rho$, while keeping $p_m \approx 1$ constant. In the right panel of Fig.~\ref{fig:12}, we consider the converse, i.e. an array of values of $\rho_m$ with $\rho \approx 1$ kept constant. Comparing the left and right panels, we are able to observe a degree of asymmetry in the values of truthfulness reached in the long run. Indeed, we see that it takes a very strong debunking effort ($\rho \rightarrow 1$) to only recover modest levels of truthfulness when disinformation is pushed hard ($\rho_m \approx 1$). At the same time, a strong disinformation campaign ($\rho_m \rightarrow 1$) is rather successful at dismantling truthfulness even when the ground truth is pushed forcefully ($\rho \approx 1$).

\begin{figure}[h!]
    \centering
      \centerline{
       \includegraphics[width=8cm]{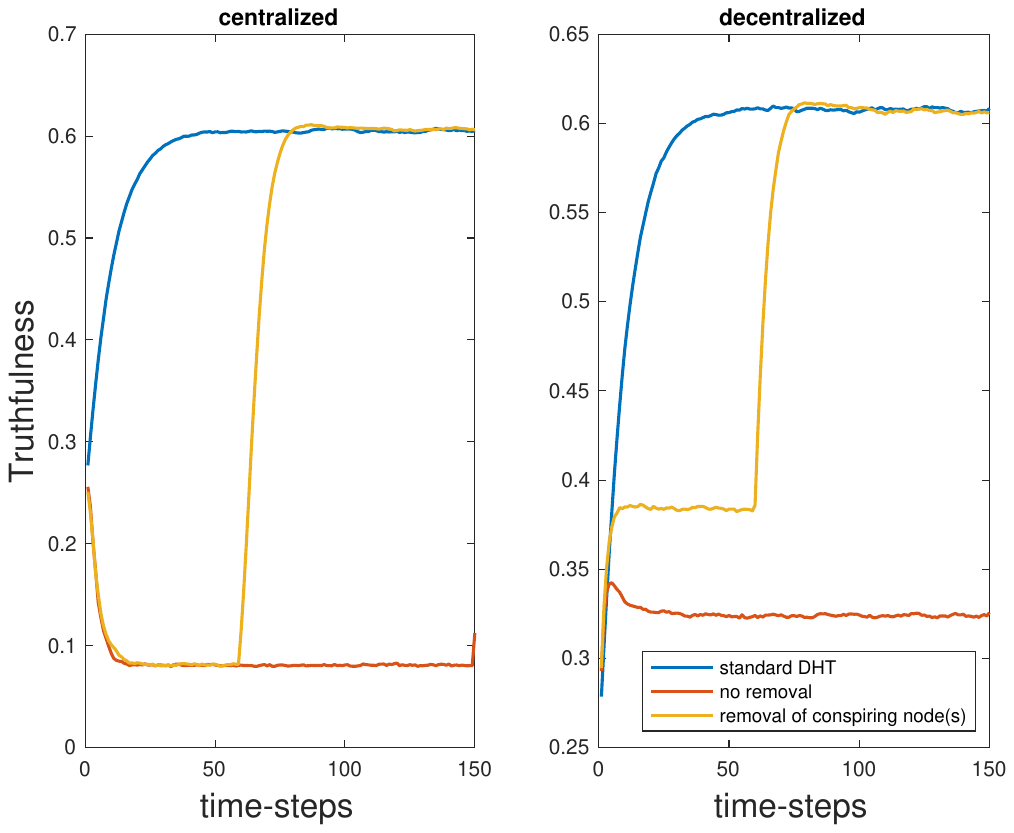}}
    \caption{ Temporal evolution of truthfulness with removal of (left) conspiring mega-node and (right) decentralized ($\beta_c=10\%$) conspirator sub-population at $60$th time-step. Results are obtained by averaging $100$ independent simulations run on sparse ER networks with $N=100$ and $M=4$. }
  
    \label{fig:13}
\end{figure}

\begin{figure}[h!]
    \centering
      \centerline{
       \includegraphics[width=5cm]{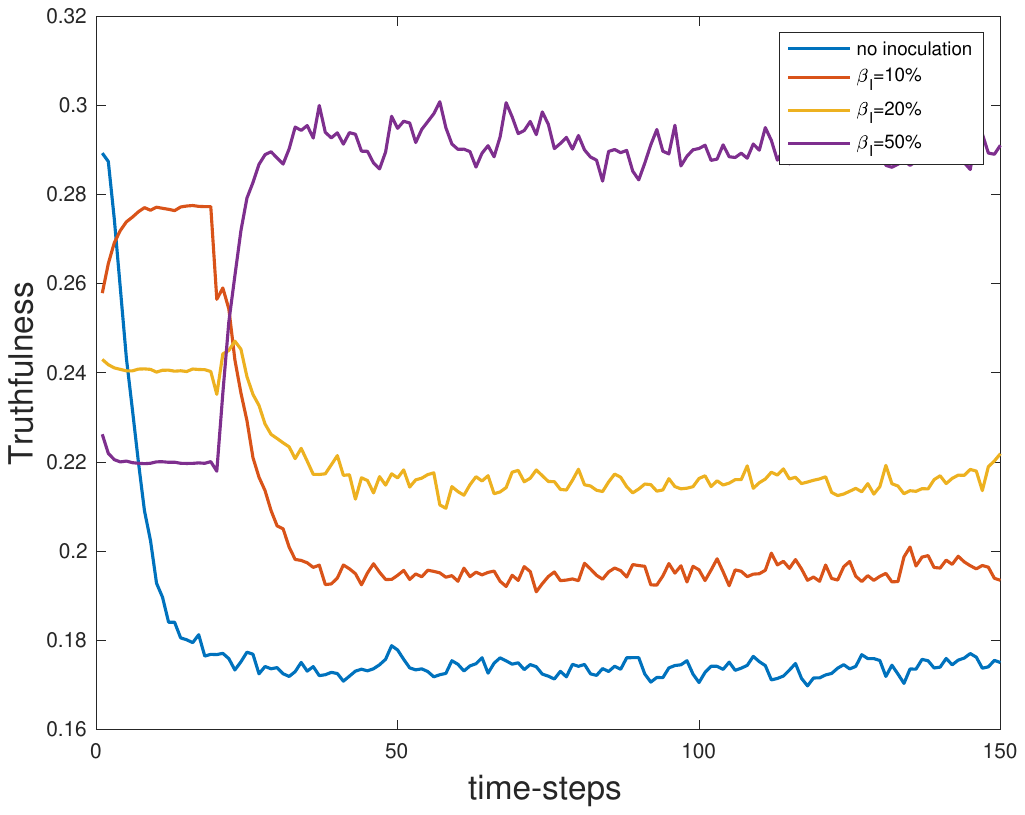}
      \includegraphics[width=5cm]{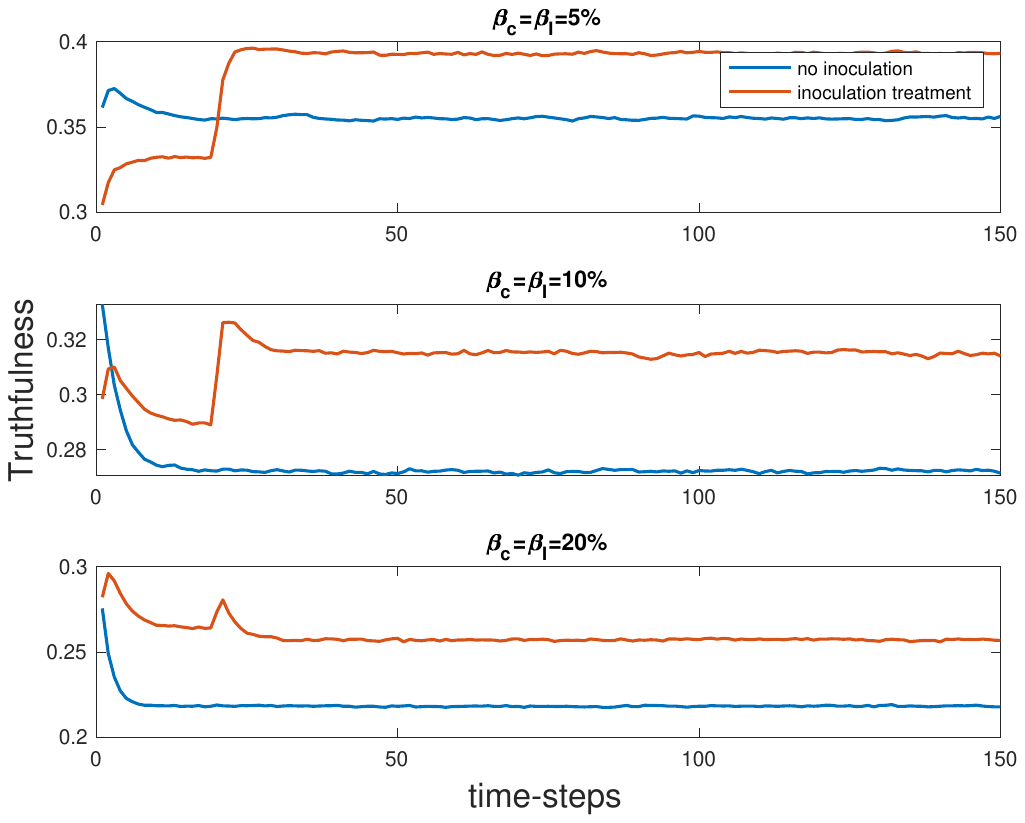}
      \includegraphics[width=5cm]{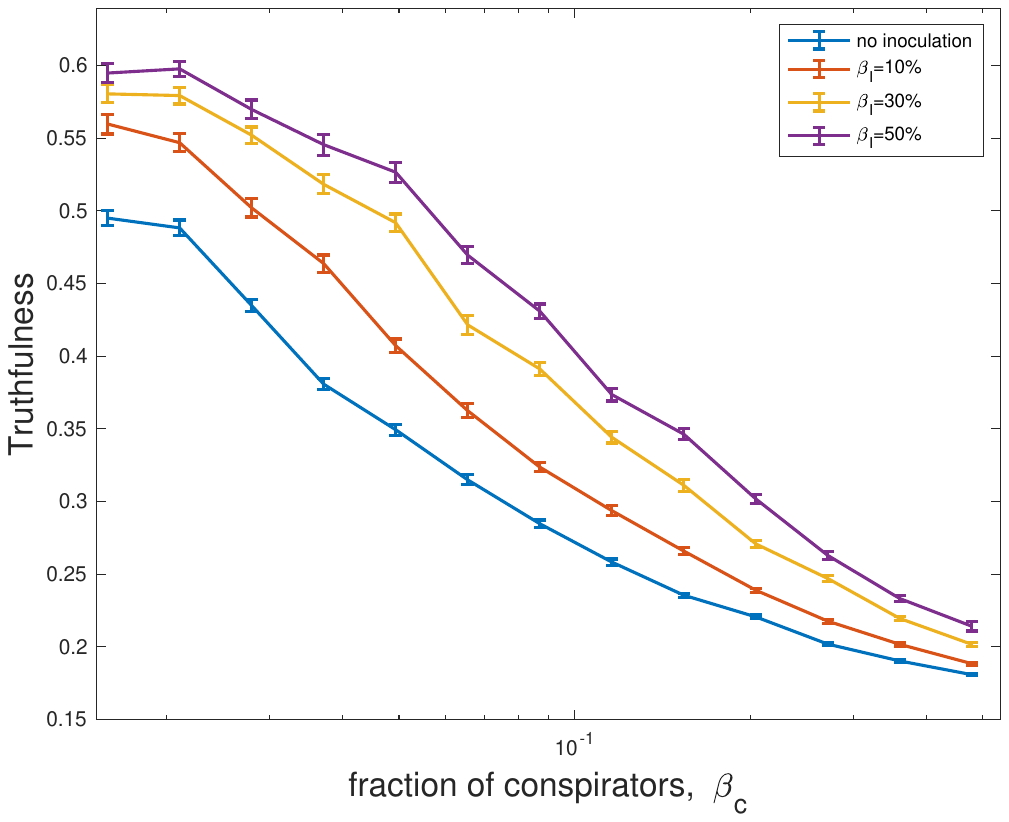}}
    \caption{Left: temporal evolution of truthfulness with a conspiring mega-node and varying fractions of decentralized mildly misinforming nodes $\beta_I$. Center: temporal evolution with decentralized conspirators and mildly misinforming nodes. Right: steady-state truthfulness as a function of the concentration of conspirator agents $\beta_c$ at different concentrations of mildly misinforming agents. In all cases we implement an inoculation phase of $T_I=20$ time steps, with $N=100$ and $M=4$ in a series of sparse BA networks.}
  
    \label{fig:14}
\end{figure}

\begin{figure}[h!]
    \centering
              \centerline{
       \includegraphics[width=7cm]{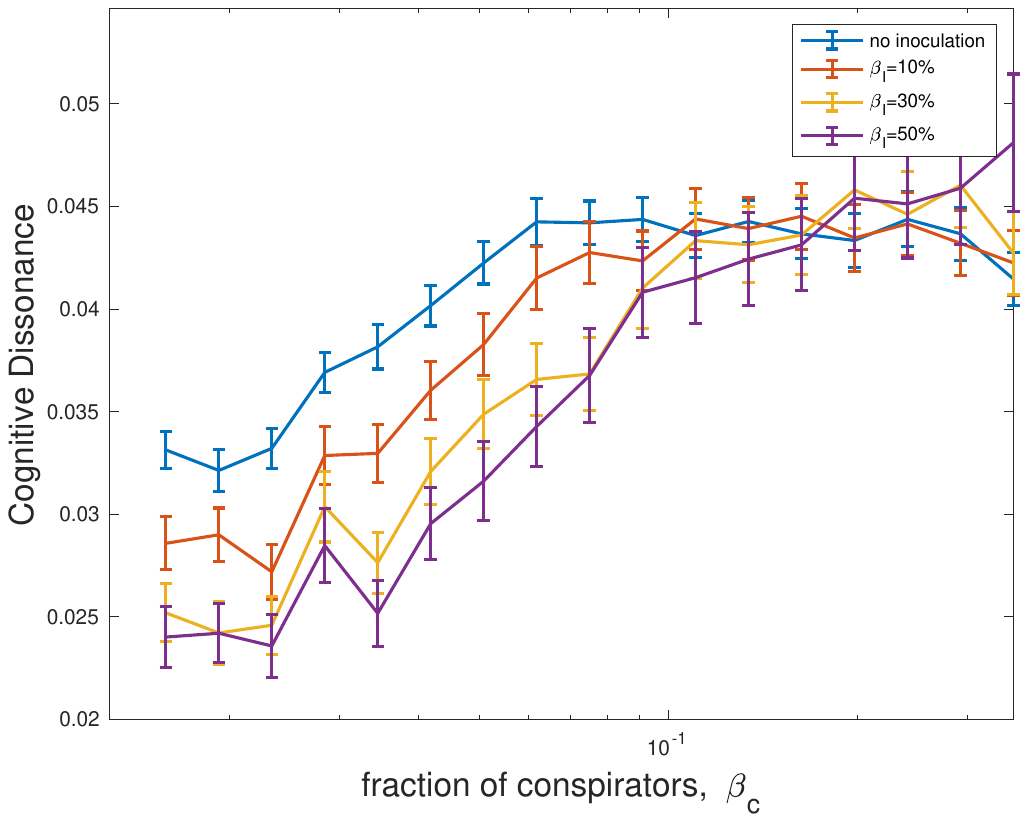}}
    \caption{Cognitive Dissonance as a function conspirator agents, $\beta_c$ at different concentrations of mildly misinforming agents with an inoculation phase of $T_I=20$ time steps, with $N=100$ and $M=4$ in  a series of sparse BA networks}
  
    \label{fig:15}
\end{figure}

We now turn to \textbf{Q5}. As mentioned, the removal or suspension of a user's account, otherwise known as \textit{deplatforming}, is common practice on social media platforms when it has been identified to have expressed or to be linked to hateful or toxic content. We simulate this policy both in a centralized and decentralized context. In the former case, we remove the conspiring mega-node, while in the latter we remove all conspirator nodes at an intermediate time-step $T_r$. We note that the network topology doesn't actually change, where `removed/suspended' nodes are not able to communicate with the network. 

In Fig.~\ref{fig:13} we may observe that, under both scenarios, once the removal of mis/dis-informing node(s) has been put in place, there is reversion to the standard DHT steady-state. While this is consistent to an extent with the literature, i.e. there is a reduction in overall attention toward such `norm-violating' users~\cite{ribeiro2024deplatforming}, this result should be interpreted with caution, given the limitations of the model due to its stylized nature. At the same time, the robustness of the DHT procedure may be highlighted from these findings where learning in a distributed setting, including within social networks, is an ongoing field of research~\cite{lalitha2018social,ntemos2021social,hare2020non}. 

Given much of mitigation techniques tend to occur post-factum, should there be further investment in actually \textit{preventing} the occurrence of misinformation? We now simulate the concept of psychological inoculation within the context of our DHT framework in an effort to resolve \textbf{Q6}. The basic premise of inoculation against misinformation, similar to the vaccine counter-part,  is to expose an individual to a weaker form of misinformation such that they would build up an immunity to it as the forms of misinformation may strengthen. Inoculation is aimed at being a preventative measure, though distinct from the intervention of prebunking. Much of the literature regarding this has been limited to the psychology literature~\cite{lewandowsky2021countering,lewandowsky2017beyond}. More specifically, inoculation games have been created, such examples including Bad News~\footnote{www.getbadnews.com} or Go Viral!~\footnote{www.goviralgame.com}, where players are given warnings which can indicate content as misinforming in some way, such that cognitive resistance may be built. Given of course the wide array of disciplines investigating the pressing issues of misinformation spread and infodemics overall, we seek to introduce and simulate this mitigation strategy to that of the opinion dynamics literature.
 
It can be realized that the inoculation process is by nature \textit{decentralized}. We introduce a fraction of mildly misinforming agents denoted by $\beta_I$ such that their public belief vector is $\boldsymbol{b}_{i}=(b_1,b_2,b_3 \ldots )$, where $b_1=0.4$, $b_i$ with $i=2 \ldots M$ determined randomly via the uniform distribution rescaled to ensure normalization and such that $\sum_k b_{i} =1$ in the sub-period of $T_I$ otherwise known as the \textit{inoculation phase}. In other words, such agents weakly push the identical agenda as that the conspiring node(s) would strongly push (which we continue to assume to be $\theta_1$ without loss of generality). Regarding topology of the network structure, we numerically simulate this within sparse BA networks. Given the existence of hubs within such networks, we select a subset of `central and influential' nodes determined via eigenvector centrality \cite{ruhnau2000eigenvector} and assign them to the condition of mildly misinforming within the inoculation phase 
 
 From $T_I$ to the final time $T$, a more strengthened iteration of misinformation enters within the network in the form of either a conspiring mega-node  (e.g. a charismatic and influential political figure) or decentralized, distributed conspiring agents  (e.g. malicious bots). In Figure \ref{fig:14}, we observe the effect of the former in the left panel and latter in the center and right panels, at different concentrations of mildly misinforming agents in the inoculation phase. Higher concentrations of $\beta_I$ appear to render higher levels of truthfulness given that the network population interacts more frequently with such mildly misinforming agents, particularly for lowers level of $\beta_c$. 
 Looking at Fig.~\ref{fig:15}, we see an agreement of sentiments in CD as a function of fraction of conspirators $\beta_c$, that is, for smaller values of $\beta_c$, we observe a noticeable diminishing of CD between networks that posses no sub-population of $\beta_I$, mildly misinforming agents and those which are `inoculated.'

\section{Conclusions} 

In this study, we aimed to investigate what entities would me most effective at combating the spread of misinformation in social networks, distinguishing between centralized and decentralized `protocols'. Through a social learning procedure (DHT) where agents aim to learn a ground-truth among a set of competing hypotheses, we observed that exposure to misinformation from centralized sources generally resulted in more vulnerable states/populations. In the same spirit, information correction via debunking originating from centralized sources tended to lead to higher levels of truthfulness. Moreover, we found prebunking to not have a significant impact in the long-run on the agents' likelihood to learn the ground-truth. More specifically, DHT agents amidst solely debunker agents appeared to be more resistant against misinformation than DHT agents to whom the ground-truth was preemptively promoted. Additionally, an asymmetry in truthfulness levels was shown to exist among forces that conspire and forces that correct, reminiscent of certain entities' political leanings or desire for expressing a `balanced' perspective.

We also simulated \textit{deplatforming}, being a practiced retroactive technique for moderating content on social media platforms. Such an implementation led the dynamics to revert to the standard DHT steady-state in truthfulness, reflecting the robustness of this particular procedure to combat misinformation. Drawing from the psychology literature, we also showed that there appeared to be positive effects (on average) in increasing the collective belief in the ground truth, more so for lower concentrations of conspirator agents, suggesting the viability of inoculation as a way to potentially build resistance against of misinformation.

For completeness, although not considered within this paper, we also make note of the calls towards content moderation and curbing of misleading or malicious information through recommender systems. Notably, recommender algorithms of social media platforms themselves present their own risks as they are known to prioritize \textit{maximum engagement}, giving rise to negativity bias~\cite{chavalarias2024can}, and an association to much of the observed toxic online behavior.

We have observed, partly in this paper, where we have presented a stylized agent-based model of opinion dynamics, in addition to considering the different literatures studying misinformation where the dedicated literature has provided various ways to classify interventions aimed at combating misinformation. In~\cite{johansson2022can}, interventions are surveyed and categorized as those that \textit{curb} and/or \textit{prepare} and/or \textit{respond}.  The classification and framing of interventions presented in this paper as \textit{centralized} versus \textit{decentralized}, and evaluated with a stylized opinion dynamics framework, may complement such efforts where such a framing as well assist from a policy stand-point. Beyond this, underlying the effectiveness of counter-strategies has been shown to be tied to indoctrination and the strength with which an individual subscribes to an ideology and the collection of beliefs which may come or be paired with an unwavering devotion~\cite{dastgeer2022qanon}. In essence, a chosen strategy may present their own set of caveats and associated consequences/risks, upon where future work may continue to progress, in addition to focusing upon context-specific efforts where such a chosen strategy may permeate. 

With that in mind, let us conclude by noting that the correcting interventions considered in this paper are stylized and are only meant as a caricature of their real-life counterparts. The advantage naturally is that this has allowed for comparisons within a simplified setting; however, one of the drawbacks associated with such a simplification is evidently the potential lack of nuance and/or of the moral dilemmas corresponding to the actuality of the circumstances faced in reality.

\bibliography{aipsamp.bib}% Produces the bibliography via BibTeX.

\providecommand{\noopsort}[1]{}\providecommand{\singleletter}[1]{#1}%
\begin{thebibliography}{41}
\expandafter\ifx\csname natexlab\endcsname\relax\def\natexlab#1{#1}\fi
\expandafter\ifx\csname bibnamefont\endcsname\relax
  \def\bibnamefont#1{#1}\fi
\expandafter\ifx\csname bibfnamefont\endcsname\relax
  \def\bibfnamefont#1{#1}\fi
\expandafter\ifx\csname citenamefont\endcsname\relax
  \def\citenamefont#1{#1}\fi
\expandafter\ifx\csname url\endcsname\relax
  \def\url#1{\texttt{#1}}\fi
\expandafter\ifx\csname urlprefix\endcsname\relax\def\urlprefix{URL }\fi
\providecommand{\bibinfo}[2]{#2}
\providecommand{\eprint}[2][]{\url{#2}}

\bibitem[{\citenamefont{Lewandowsky et~al.}(2017)\citenamefont{Lewandowsky,
  Ecker, and Cook}}]{lewandowsky2017beyond}
\bibinfo{author}{\bibfnamefont{S.}~\bibnamefont{Lewandowsky}},
  \bibinfo{author}{\bibfnamefont{U.~K.} \bibnamefont{Ecker}}, \bibnamefont{and}
  \bibinfo{author}{\bibfnamefont{J.}~\bibnamefont{Cook}},
  \bibinfo{journal}{Journal of applied research in memory and cognition}
  \textbf{\bibinfo{volume}{6}}, \bibinfo{pages}{353} (\bibinfo{year}{2017}).

\bibitem[{\citenamefont{Spohr}(2017)}]{spohr2017fake}
\bibinfo{author}{\bibfnamefont{D.}~\bibnamefont{Spohr}},
  \bibinfo{journal}{Business information review} \textbf{\bibinfo{volume}{34}},
  \bibinfo{pages}{150} (\bibinfo{year}{2017}).

\bibitem[{\citenamefont{Gallotti et~al.}(2020)\citenamefont{Gallotti, Valle,
  Castaldo, Sacco, and De~Domenico}}]{gallotti2020assessing}
\bibinfo{author}{\bibfnamefont{R.}~\bibnamefont{Gallotti}},
  \bibinfo{author}{\bibfnamefont{F.}~\bibnamefont{Valle}},
  \bibinfo{author}{\bibfnamefont{N.}~\bibnamefont{Castaldo}},
  \bibinfo{author}{\bibfnamefont{P.}~\bibnamefont{Sacco}}, \bibnamefont{and}
  \bibinfo{author}{\bibfnamefont{M.}~\bibnamefont{De~Domenico}},
  \bibinfo{journal}{Nature human behaviour} \textbf{\bibinfo{volume}{4}},
  \bibinfo{pages}{1285} (\bibinfo{year}{2020}).

\bibitem[{\citenamefont{Hornik et~al.}(2015)\citenamefont{Hornik, Satchi,
  Cesareo, and Pastore}}]{hornik2015information}
\bibinfo{author}{\bibfnamefont{J.}~\bibnamefont{Hornik}},
  \bibinfo{author}{\bibfnamefont{R.~S.} \bibnamefont{Satchi}},
  \bibinfo{author}{\bibfnamefont{L.}~\bibnamefont{Cesareo}}, \bibnamefont{and}
  \bibinfo{author}{\bibfnamefont{A.}~\bibnamefont{Pastore}},
  \bibinfo{journal}{Computers in Human Behavior} \textbf{\bibinfo{volume}{45}},
  \bibinfo{pages}{273} (\bibinfo{year}{2015}).

\bibitem[{\citenamefont{Prior}(2003)}]{prior2003any}
\bibinfo{author}{\bibfnamefont{M.}~\bibnamefont{Prior}},
  \bibinfo{journal}{Political communication} \textbf{\bibinfo{volume}{20}},
  \bibinfo{pages}{149} (\bibinfo{year}{2003}).

\bibitem[{\citenamefont{Zhang and Ghorbani}(2020)}]{zhang2020overview}
\bibinfo{author}{\bibfnamefont{X.}~\bibnamefont{Zhang}} \bibnamefont{and}
  \bibinfo{author}{\bibfnamefont{A.~A.} \bibnamefont{Ghorbani}},
  \bibinfo{journal}{Information Processing \& Management}
  \textbf{\bibinfo{volume}{57}}, \bibinfo{pages}{102025}
  (\bibinfo{year}{2020}).

\bibitem[{\citenamefont{Duguay}(2018)}]{duguay2018social}
\bibinfo{author}{\bibfnamefont{S.}~\bibnamefont{Duguay}},
  \bibinfo{journal}{Media International Australia}
  \textbf{\bibinfo{volume}{166}}, \bibinfo{pages}{20} (\bibinfo{year}{2018}).

\bibitem[{\citenamefont{Helfers and Ebersbach}(2023)}]{helfers2023differential}
\bibinfo{author}{\bibfnamefont{A.}~\bibnamefont{Helfers}} \bibnamefont{and}
  \bibinfo{author}{\bibfnamefont{M.}~\bibnamefont{Ebersbach}},
  \bibinfo{journal}{Journal of Communication in Healthcare}
  \textbf{\bibinfo{volume}{16}}, \bibinfo{pages}{113} (\bibinfo{year}{2023}).

\bibitem[{\citenamefont{Zollo et~al.}(2017)\citenamefont{Zollo, Bessi,
  Del~Vicario, Scala, Caldarelli, Shekhtman, Havlin, and
  Quattrociocchi}}]{zollo2017debunking}
\bibinfo{author}{\bibfnamefont{F.}~\bibnamefont{Zollo}},
  \bibinfo{author}{\bibfnamefont{A.}~\bibnamefont{Bessi}},
  \bibinfo{author}{\bibfnamefont{M.}~\bibnamefont{Del~Vicario}},
  \bibinfo{author}{\bibfnamefont{A.}~\bibnamefont{Scala}},
  \bibinfo{author}{\bibfnamefont{G.}~\bibnamefont{Caldarelli}},
  \bibinfo{author}{\bibfnamefont{L.}~\bibnamefont{Shekhtman}},
  \bibinfo{author}{\bibfnamefont{S.}~\bibnamefont{Havlin}}, \bibnamefont{and}
  \bibinfo{author}{\bibfnamefont{W.}~\bibnamefont{Quattrociocchi}},
  \bibinfo{journal}{PloS one} \textbf{\bibinfo{volume}{12}},
  \bibinfo{pages}{e0181821} (\bibinfo{year}{2017}).

\bibitem[{\citenamefont{Ecker et~al.}(2011)\citenamefont{Ecker, Lewandowsky,
  Swire, and Chang}}]{ecker2011correcting}
\bibinfo{author}{\bibfnamefont{U.~K.} \bibnamefont{Ecker}},
  \bibinfo{author}{\bibfnamefont{S.}~\bibnamefont{Lewandowsky}},
  \bibinfo{author}{\bibfnamefont{B.}~\bibnamefont{Swire}}, \bibnamefont{and}
  \bibinfo{author}{\bibfnamefont{D.}~\bibnamefont{Chang}},
  \bibinfo{journal}{Psychonomic bulletin \& review}
  \textbf{\bibinfo{volume}{18}}, \bibinfo{pages}{570} (\bibinfo{year}{2011}).

\bibitem[{\citenamefont{Vinck et~al.}(2019)\citenamefont{Vinck, Pham, Bindu,
  Bedford, and Nilles}}]{vinck2019institutional}
\bibinfo{author}{\bibfnamefont{P.}~\bibnamefont{Vinck}},
  \bibinfo{author}{\bibfnamefont{P.~N.} \bibnamefont{Pham}},
  \bibinfo{author}{\bibfnamefont{K.~K.} \bibnamefont{Bindu}},
  \bibinfo{author}{\bibfnamefont{J.}~\bibnamefont{Bedford}}, \bibnamefont{and}
  \bibinfo{author}{\bibfnamefont{E.~J.} \bibnamefont{Nilles}},
  \bibinfo{journal}{The Lancet Infectious Diseases}
  \textbf{\bibinfo{volume}{19}}, \bibinfo{pages}{529} (\bibinfo{year}{2019}).

\bibitem[{\citenamefont{Xiao et~al.}(2021)\citenamefont{Xiao, Borah, and
  Su}}]{xiao2021dangers}
\bibinfo{author}{\bibfnamefont{X.}~\bibnamefont{Xiao}},
  \bibinfo{author}{\bibfnamefont{P.}~\bibnamefont{Borah}}, \bibnamefont{and}
  \bibinfo{author}{\bibfnamefont{Y.}~\bibnamefont{Su}},
  \bibinfo{journal}{Public Understanding of Science}
  \textbf{\bibinfo{volume}{30}}, \bibinfo{pages}{977} (\bibinfo{year}{2021}).

\bibitem[{\citenamefont{Jang et~al.}(2019)\citenamefont{Jang, Lee, and
  Shin}}]{jang2019debunking}
\bibinfo{author}{\bibfnamefont{J.-w.} \bibnamefont{Jang}},
  \bibinfo{author}{\bibfnamefont{E.-J.} \bibnamefont{Lee}}, \bibnamefont{and}
  \bibinfo{author}{\bibfnamefont{S.~Y.} \bibnamefont{Shin}},
  \bibinfo{journal}{Cyberpsychology, Behavior, and Social Networking}
  \textbf{\bibinfo{volume}{22}}, \bibinfo{pages}{423} (\bibinfo{year}{2019}).

\bibitem[{\citenamefont{Jhaver et~al.}(2021)\citenamefont{Jhaver, Boylston,
  Yang, and Bruckman}}]{jhaver2021evaluating}
\bibinfo{author}{\bibfnamefont{S.}~\bibnamefont{Jhaver}},
  \bibinfo{author}{\bibfnamefont{C.}~\bibnamefont{Boylston}},
  \bibinfo{author}{\bibfnamefont{D.}~\bibnamefont{Yang}}, \bibnamefont{and}
  \bibinfo{author}{\bibfnamefont{A.}~\bibnamefont{Bruckman}},
  \bibinfo{journal}{Proceedings of the ACM on human-computer interaction}
  \textbf{\bibinfo{volume}{5}}, \bibinfo{pages}{1} (\bibinfo{year}{2021}).

\bibitem[{\citenamefont{Klinenberg}(2024)}]{klinenberg2024does}
\bibinfo{author}{\bibfnamefont{D.}~\bibnamefont{Klinenberg}},
  \bibinfo{journal}{Journal of Conflict Resolution}
  \textbf{\bibinfo{volume}{68}}, \bibinfo{pages}{1199} (\bibinfo{year}{2024}).

\bibitem[{\citenamefont{Ali et~al.}(2021)\citenamefont{Ali, Saeed, Aldreabi,
  Blackburn, De~Cristofaro, Zannettou, and Stringhini}}]{ali2021understanding}
\bibinfo{author}{\bibfnamefont{S.}~\bibnamefont{Ali}},
  \bibinfo{author}{\bibfnamefont{M.~H.} \bibnamefont{Saeed}},
  \bibinfo{author}{\bibfnamefont{E.}~\bibnamefont{Aldreabi}},
  \bibinfo{author}{\bibfnamefont{J.}~\bibnamefont{Blackburn}},
  \bibinfo{author}{\bibfnamefont{E.}~\bibnamefont{De~Cristofaro}},
  \bibinfo{author}{\bibfnamefont{S.}~\bibnamefont{Zannettou}},
  \bibnamefont{and}
  \bibinfo{author}{\bibfnamefont{G.}~\bibnamefont{Stringhini}}, in
  \emph{\bibinfo{booktitle}{Proceedings of the 13th ACM Web Science Conference
  2021}} (\bibinfo{year}{2021}), pp. \bibinfo{pages}{187--195}.

\bibitem[{\citenamefont{Lewandowsky and Van
  Der~Linden}(2021)}]{lewandowsky2021countering}
\bibinfo{author}{\bibfnamefont{S.}~\bibnamefont{Lewandowsky}} \bibnamefont{and}
  \bibinfo{author}{\bibfnamefont{S.}~\bibnamefont{Van Der~Linden}},
  \bibinfo{journal}{European Review of Social Psychology}
  \textbf{\bibinfo{volume}{32}}, \bibinfo{pages}{348} (\bibinfo{year}{2021}).

\bibitem[{\citenamefont{Cook et~al.}(2017)\citenamefont{Cook, Lewandowsky, and
  Ecker}}]{cook2017neutralizing}
\bibinfo{author}{\bibfnamefont{J.}~\bibnamefont{Cook}},
  \bibinfo{author}{\bibfnamefont{S.}~\bibnamefont{Lewandowsky}},
  \bibnamefont{and} \bibinfo{author}{\bibfnamefont{U.~K.} \bibnamefont{Ecker}},
  \bibinfo{journal}{PloS one} \textbf{\bibinfo{volume}{12}},
  \bibinfo{pages}{e0175799} (\bibinfo{year}{2017}).

\bibitem[{\citenamefont{Lalitha et~al.}(2018)\citenamefont{Lalitha, Javidi, and
  Sarwate}}]{lalitha2018social}
\bibinfo{author}{\bibfnamefont{A.}~\bibnamefont{Lalitha}},
  \bibinfo{author}{\bibfnamefont{T.}~\bibnamefont{Javidi}}, \bibnamefont{and}
  \bibinfo{author}{\bibfnamefont{A.~D.} \bibnamefont{Sarwate}},
  \bibinfo{journal}{IEEE Transactions on Information Theory}
  \textbf{\bibinfo{volume}{64}}, \bibinfo{pages}{6161} (\bibinfo{year}{2018}).

\bibitem[{\citenamefont{Riazi and Livan}(2024{\natexlab{a}})}]{riazi2024public}
\bibinfo{author}{\bibfnamefont{D.}~\bibnamefont{Riazi}} \bibnamefont{and}
  \bibinfo{author}{\bibfnamefont{G.}~\bibnamefont{Livan}},
  \bibinfo{journal}{Physica A: Statistical Mechanics and its Applications} p.
  \bibinfo{pages}{129621} (\bibinfo{year}{2024}{\natexlab{a}}).

\bibitem[{\citenamefont{Riazi and
  Livan}(2024{\natexlab{b}})}]{riazi2024mitigating}
\bibinfo{author}{\bibfnamefont{D.}~\bibnamefont{Riazi}} \bibnamefont{and}
  \bibinfo{author}{\bibfnamefont{G.}~\bibnamefont{Livan}},
  \bibinfo{journal}{arXiv preprint arXiv:2403.13630}
  (\bibinfo{year}{2024}{\natexlab{b}}).

\bibitem[{\citenamefont{Cooper}(2019)}]{cooper2019cognitive}
\bibinfo{author}{\bibfnamefont{J.}~\bibnamefont{Cooper}},
  \bibinfo{journal}{International Review of Social Psychology}
  \textbf{\bibinfo{volume}{32}}, \bibinfo{pages}{7} (\bibinfo{year}{2019}).

\bibitem[{\citenamefont{Stern and Livan}(2021)}]{stern2021impact}
\bibinfo{author}{\bibfnamefont{S.}~\bibnamefont{Stern}} \bibnamefont{and}
  \bibinfo{author}{\bibfnamefont{G.}~\bibnamefont{Livan}},
  \bibinfo{journal}{Royal Society open science} \textbf{\bibinfo{volume}{8}},
  \bibinfo{pages}{201943} (\bibinfo{year}{2021}).

\bibitem[{\citenamefont{Davidson}(2017)}]{davidson2017vaccination}
\bibinfo{author}{\bibfnamefont{M.}~\bibnamefont{Davidson}},
  \bibinfo{journal}{Dialogues in clinical neuroscience}
  \textbf{\bibinfo{volume}{19}}, \bibinfo{pages}{403} (\bibinfo{year}{2017}).

\bibitem[{\citenamefont{Sikder et~al.}(2020)\citenamefont{Sikder, Smith, Vivo,
  and Livan}}]{sikder2020minimalistic}
\bibinfo{author}{\bibfnamefont{O.}~\bibnamefont{Sikder}},
  \bibinfo{author}{\bibfnamefont{R.~E.} \bibnamefont{Smith}},
  \bibinfo{author}{\bibfnamefont{P.}~\bibnamefont{Vivo}}, \bibnamefont{and}
  \bibinfo{author}{\bibfnamefont{G.}~\bibnamefont{Livan}},
  \bibinfo{journal}{Scientific reports} \textbf{\bibinfo{volume}{10}},
  \bibinfo{pages}{5493} (\bibinfo{year}{2020}).

\bibitem[{\citenamefont{Ecker et~al.}(2022)\citenamefont{Ecker, Lewandowsky,
  Cook, Schmid, Fazio, Brashier, Kendeou, Vraga, and
  Amazeen}}]{ecker2022psychological}
\bibinfo{author}{\bibfnamefont{U.~K.} \bibnamefont{Ecker}},
  \bibinfo{author}{\bibfnamefont{S.}~\bibnamefont{Lewandowsky}},
  \bibinfo{author}{\bibfnamefont{J.}~\bibnamefont{Cook}},
  \bibinfo{author}{\bibfnamefont{P.}~\bibnamefont{Schmid}},
  \bibinfo{author}{\bibfnamefont{L.~K.} \bibnamefont{Fazio}},
  \bibinfo{author}{\bibfnamefont{N.}~\bibnamefont{Brashier}},
  \bibinfo{author}{\bibfnamefont{P.}~\bibnamefont{Kendeou}},
  \bibinfo{author}{\bibfnamefont{E.~K.} \bibnamefont{Vraga}}, \bibnamefont{and}
  \bibinfo{author}{\bibfnamefont{M.~A.} \bibnamefont{Amazeen}},
  \bibinfo{journal}{Nature Reviews Psychology} \textbf{\bibinfo{volume}{1}},
  \bibinfo{pages}{13} (\bibinfo{year}{2022}).

\bibitem[{\citenamefont{Vegetti}(2019)}]{vegetti2019political}
\bibinfo{author}{\bibfnamefont{F.}~\bibnamefont{Vegetti}},
  \bibinfo{journal}{The ANNALS of the American Academy of Political and Social
  Science} \textbf{\bibinfo{volume}{681}}, \bibinfo{pages}{78}
  (\bibinfo{year}{2019}).

\bibitem[{\citenamefont{Barab{\'a}si and Albert}(1999)}]{barabasi1999emergence}
\bibinfo{author}{\bibfnamefont{A.-L.} \bibnamefont{Barab{\'a}si}}
  \bibnamefont{and} \bibinfo{author}{\bibfnamefont{R.}~\bibnamefont{Albert}},
  \bibinfo{journal}{science} \textbf{\bibinfo{volume}{286}},
  \bibinfo{pages}{509} (\bibinfo{year}{1999}).

\bibitem[{\citenamefont{Tay et~al.}(2022)\citenamefont{Tay, Hurlstone, Kurz,
  and Ecker}}]{tay2022comparison}
\bibinfo{author}{\bibfnamefont{L.~Q.} \bibnamefont{Tay}},
  \bibinfo{author}{\bibfnamefont{M.~J.} \bibnamefont{Hurlstone}},
  \bibinfo{author}{\bibfnamefont{T.}~\bibnamefont{Kurz}}, \bibnamefont{and}
  \bibinfo{author}{\bibfnamefont{U.~K.} \bibnamefont{Ecker}},
  \bibinfo{journal}{British Journal of Psychology}
  \textbf{\bibinfo{volume}{113}}, \bibinfo{pages}{591} (\bibinfo{year}{2022}).

\bibitem[{\citenamefont{Jolley and Douglas}(2017)}]{jolley2017prevention}
\bibinfo{author}{\bibfnamefont{D.}~\bibnamefont{Jolley}} \bibnamefont{and}
  \bibinfo{author}{\bibfnamefont{K.~M.} \bibnamefont{Douglas}},
  \bibinfo{journal}{Journal of Applied Social Psychology}
  \textbf{\bibinfo{volume}{47}}, \bibinfo{pages}{459} (\bibinfo{year}{2017}).

\bibitem[{\citenamefont{Barab{\'a}si and Bonabeau}(2003)}]{barabasi2003scale}
\bibinfo{author}{\bibfnamefont{A.-L.} \bibnamefont{Barab{\'a}si}}
  \bibnamefont{and} \bibinfo{author}{\bibfnamefont{E.}~\bibnamefont{Bonabeau}},
  \bibinfo{journal}{Scientific american} \textbf{\bibinfo{volume}{288}},
  \bibinfo{pages}{60} (\bibinfo{year}{2003}).

\bibitem[{\citenamefont{McCright and
  Dunlap}(2011)}]{mccright2011politicization}
\bibinfo{author}{\bibfnamefont{A.~M.} \bibnamefont{McCright}} \bibnamefont{and}
  \bibinfo{author}{\bibfnamefont{R.~E.} \bibnamefont{Dunlap}},
  \bibinfo{journal}{The Sociological Quarterly} \textbf{\bibinfo{volume}{52}},
  \bibinfo{pages}{155} (\bibinfo{year}{2011}).

\bibitem[{\citenamefont{Wang et~al.}(2019)\citenamefont{Wang, McKee, Torbica,
  and Stuckler}}]{wang2019systematic}
\bibinfo{author}{\bibfnamefont{Y.}~\bibnamefont{Wang}},
  \bibinfo{author}{\bibfnamefont{M.}~\bibnamefont{McKee}},
  \bibinfo{author}{\bibfnamefont{A.}~\bibnamefont{Torbica}}, \bibnamefont{and}
  \bibinfo{author}{\bibfnamefont{D.}~\bibnamefont{Stuckler}},
  \bibinfo{journal}{Social science \& medicine} \textbf{\bibinfo{volume}{240}},
  \bibinfo{pages}{112552} (\bibinfo{year}{2019}).

\bibitem[{\citenamefont{Lewandowsky et~al.}(2012)\citenamefont{Lewandowsky,
  Ecker, Seifert, Schwarz, and Cook}}]{lewandowsky2012misinformation}
\bibinfo{author}{\bibfnamefont{S.}~\bibnamefont{Lewandowsky}},
  \bibinfo{author}{\bibfnamefont{U.~K.} \bibnamefont{Ecker}},
  \bibinfo{author}{\bibfnamefont{C.~M.} \bibnamefont{Seifert}},
  \bibinfo{author}{\bibfnamefont{N.}~\bibnamefont{Schwarz}}, \bibnamefont{and}
  \bibinfo{author}{\bibfnamefont{J.}~\bibnamefont{Cook}},
  \bibinfo{journal}{Psychological science in the public interest}
  \textbf{\bibinfo{volume}{13}}, \bibinfo{pages}{106} (\bibinfo{year}{2012}).

\bibitem[{\citenamefont{Ribeiro et~al.}(2024)\citenamefont{Ribeiro, Jhaver,
  Reignier-Tayar, West et~al.}}]{ribeiro2024deplatforming}
\bibinfo{author}{\bibfnamefont{M.~H.} \bibnamefont{Ribeiro}},
  \bibinfo{author}{\bibfnamefont{S.}~\bibnamefont{Jhaver}},
  \bibinfo{author}{\bibfnamefont{M.}~\bibnamefont{Reignier-Tayar}},
  \bibinfo{author}{\bibfnamefont{R.}~\bibnamefont{West}}, \bibnamefont{et~al.},
  \bibinfo{journal}{arXiv preprint arXiv:2401.01253}  (\bibinfo{year}{2024}).

\bibitem[{\citenamefont{Ntemos et~al.}(2021)\citenamefont{Ntemos, Bordignon,
  Vlaski, and Sayed}}]{ntemos2021social}
\bibinfo{author}{\bibfnamefont{K.}~\bibnamefont{Ntemos}},
  \bibinfo{author}{\bibfnamefont{V.}~\bibnamefont{Bordignon}},
  \bibinfo{author}{\bibfnamefont{S.}~\bibnamefont{Vlaski}}, \bibnamefont{and}
  \bibinfo{author}{\bibfnamefont{A.~H.} \bibnamefont{Sayed}}, in
  \emph{\bibinfo{booktitle}{ICASSP 2021-2021 IEEE International Conference on
  Acoustics, Speech and Signal Processing (ICASSP)}}
  (\bibinfo{organization}{IEEE}, \bibinfo{year}{2021}), pp.
  \bibinfo{pages}{5479--5483}.

\bibitem[{\citenamefont{Hare et~al.}(2020)\citenamefont{Hare, Uribe, Kaplan,
  and Jadbabaie}}]{hare2020non}
\bibinfo{author}{\bibfnamefont{J.~Z.} \bibnamefont{Hare}},
  \bibinfo{author}{\bibfnamefont{C.~A.} \bibnamefont{Uribe}},
  \bibinfo{author}{\bibfnamefont{L.}~\bibnamefont{Kaplan}}, \bibnamefont{and}
  \bibinfo{author}{\bibfnamefont{A.}~\bibnamefont{Jadbabaie}},
  \bibinfo{journal}{IEEE Transactions on Signal Processing}
  \textbf{\bibinfo{volume}{68}}, \bibinfo{pages}{4178} (\bibinfo{year}{2020}).

\bibitem[{\citenamefont{Ruhnau}(2000)}]{ruhnau2000eigenvector}
\bibinfo{author}{\bibfnamefont{B.}~\bibnamefont{Ruhnau}},
  \bibinfo{journal}{Social networks} \textbf{\bibinfo{volume}{22}},
  \bibinfo{pages}{357} (\bibinfo{year}{2000}).

\bibitem[{\citenamefont{Chavalarias et~al.}(2024)\citenamefont{Chavalarias,
  Bouchaud, and Panahi}}]{chavalarias2024can}
\bibinfo{author}{\bibfnamefont{D.}~\bibnamefont{Chavalarias}},
  \bibinfo{author}{\bibfnamefont{P.}~\bibnamefont{Bouchaud}}, \bibnamefont{and}
  \bibinfo{author}{\bibfnamefont{M.}~\bibnamefont{Panahi}},
  \bibinfo{journal}{Journal of Artificial Societies and Social Simulation}
  \textbf{\bibinfo{volume}{27}}, \bibinfo{pages}{1} (\bibinfo{year}{2024}).

\bibitem[{\citenamefont{Johansson et~al.}(2022)\citenamefont{Johansson, Enock,
  Hale, Vidgen, Bereskin, Margetts, and Bright}}]{johansson2022can}
\bibinfo{author}{\bibfnamefont{P.}~\bibnamefont{Johansson}},
  \bibinfo{author}{\bibfnamefont{F.}~\bibnamefont{Enock}},
  \bibinfo{author}{\bibfnamefont{S.}~\bibnamefont{Hale}},
  \bibinfo{author}{\bibfnamefont{B.}~\bibnamefont{Vidgen}},
  \bibinfo{author}{\bibfnamefont{C.}~\bibnamefont{Bereskin}},
  \bibinfo{author}{\bibfnamefont{H.}~\bibnamefont{Margetts}}, \bibnamefont{and}
  \bibinfo{author}{\bibfnamefont{J.}~\bibnamefont{Bright}},
  \bibinfo{journal}{arXiv preprint arXiv:2212.11864}  (\bibinfo{year}{2022}).

\bibitem[{\citenamefont{Dastgeer and Thapaliya}(2022)}]{dastgeer2022qanon}
\bibinfo{author}{\bibfnamefont{S.}~\bibnamefont{Dastgeer}} \bibnamefont{and}
  \bibinfo{author}{\bibfnamefont{R.}~\bibnamefont{Thapaliya}}, in
  \emph{\bibinfo{booktitle}{The Emerald Handbook of Computer-Mediated
  Communication and Social Media}} (\bibinfo{publisher}{Emerald Publishing
  Limited}, \bibinfo{year}{2022}), pp. \bibinfo{pages}{251--268}.

\end{thebibliography}

\newpage

\appendix

\section{Appendix}
\begin{figure}[h!]
    \centering
      \centerline{
      \includegraphics[width=7cm]{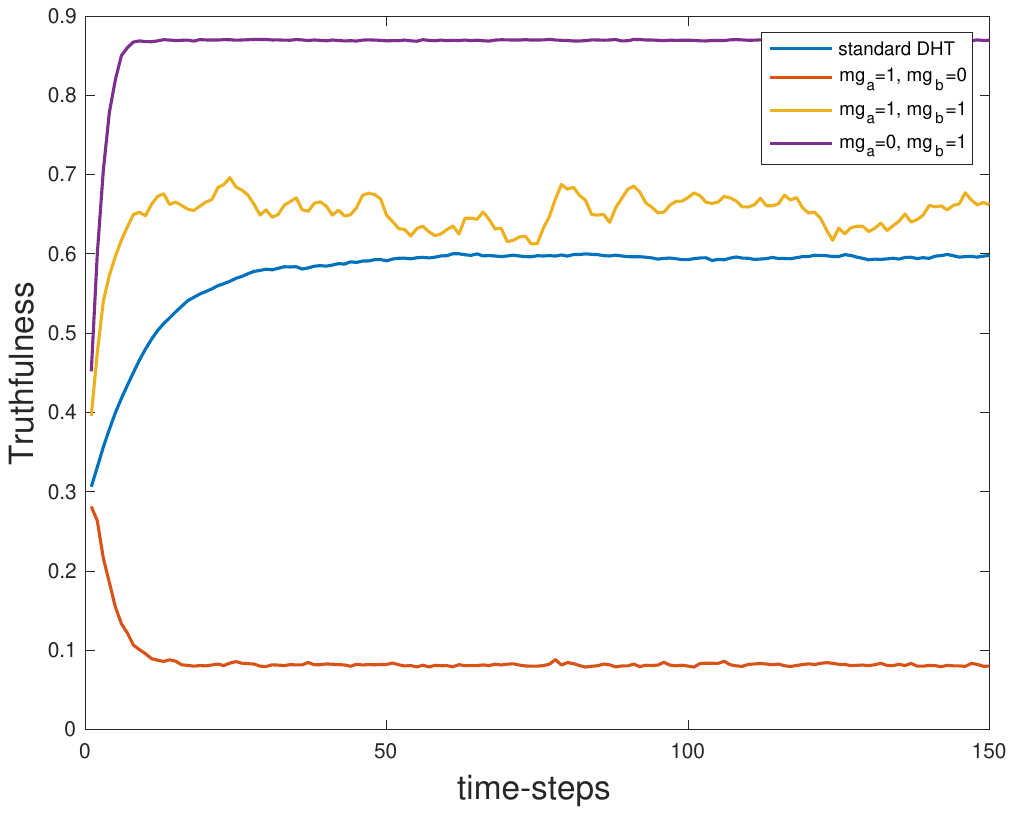}
      \includegraphics[width=7cm]{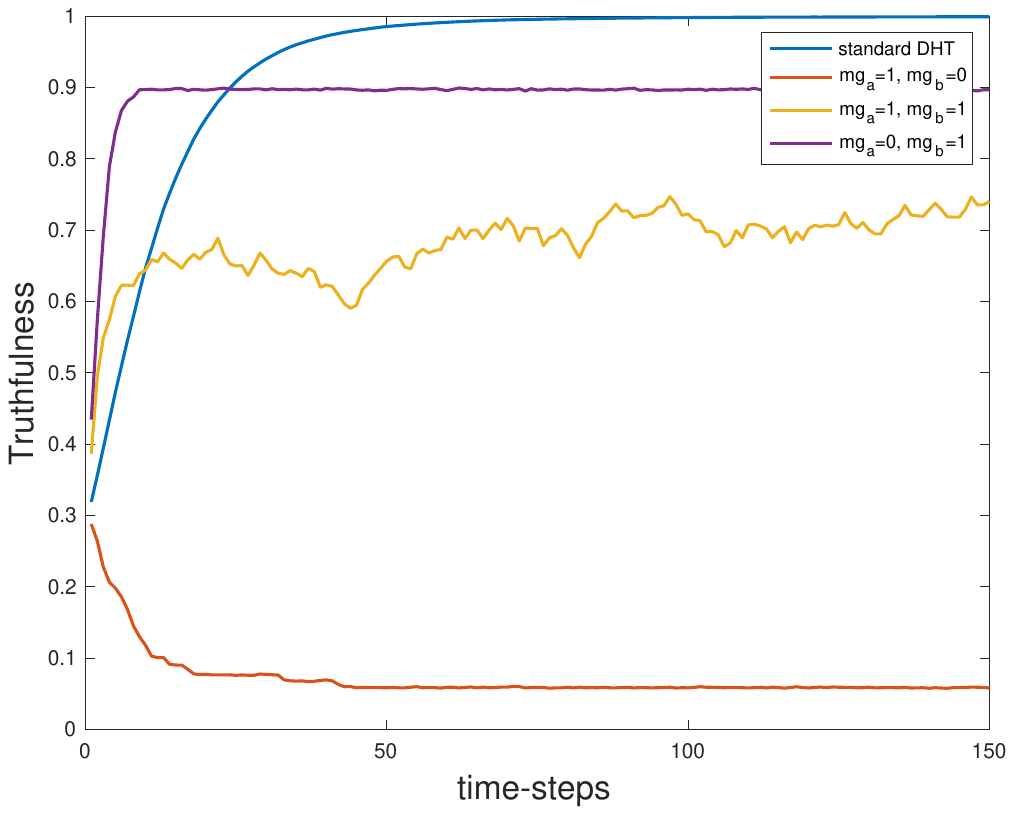}}
    \caption{Truthfulness under `mega-nodes' in a series of BA networks (left) $k=10$, (right) $k=100$ with $N=100$ and $M=4$}
    \label{fig:9}
\end{figure}

\begin{figure}[h!]
    \centering
      \centerline{
      \includegraphics[width=7cm]{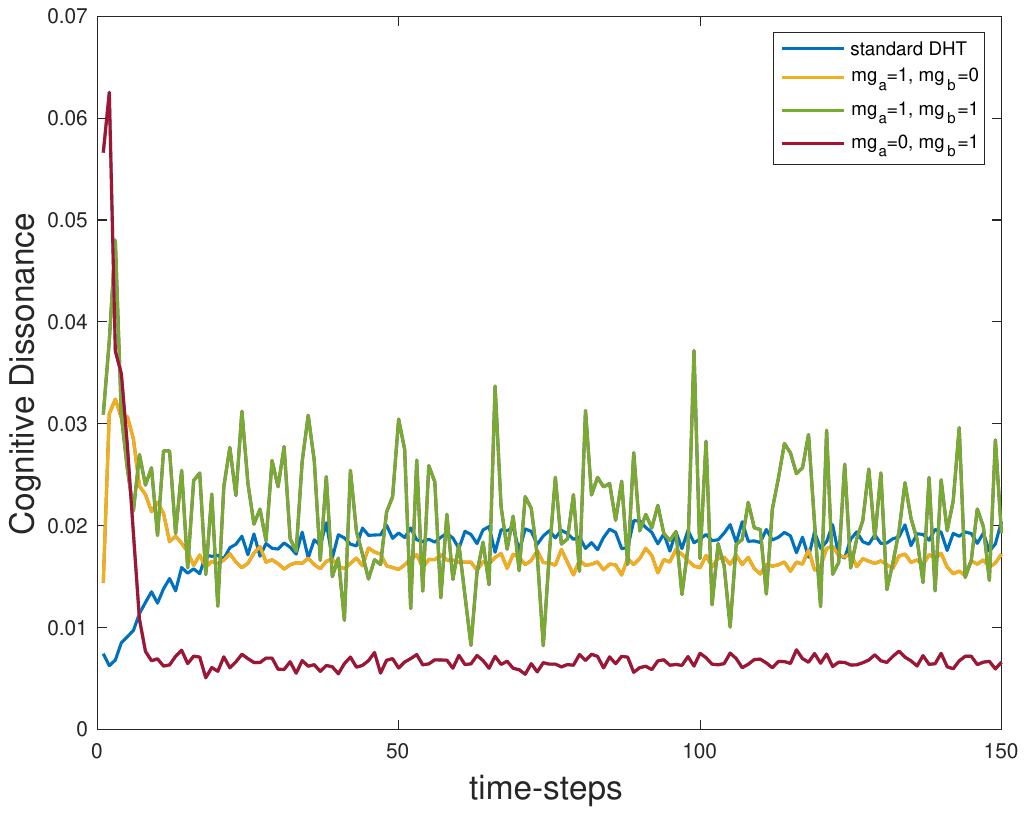}
      \includegraphics[width=7cm]{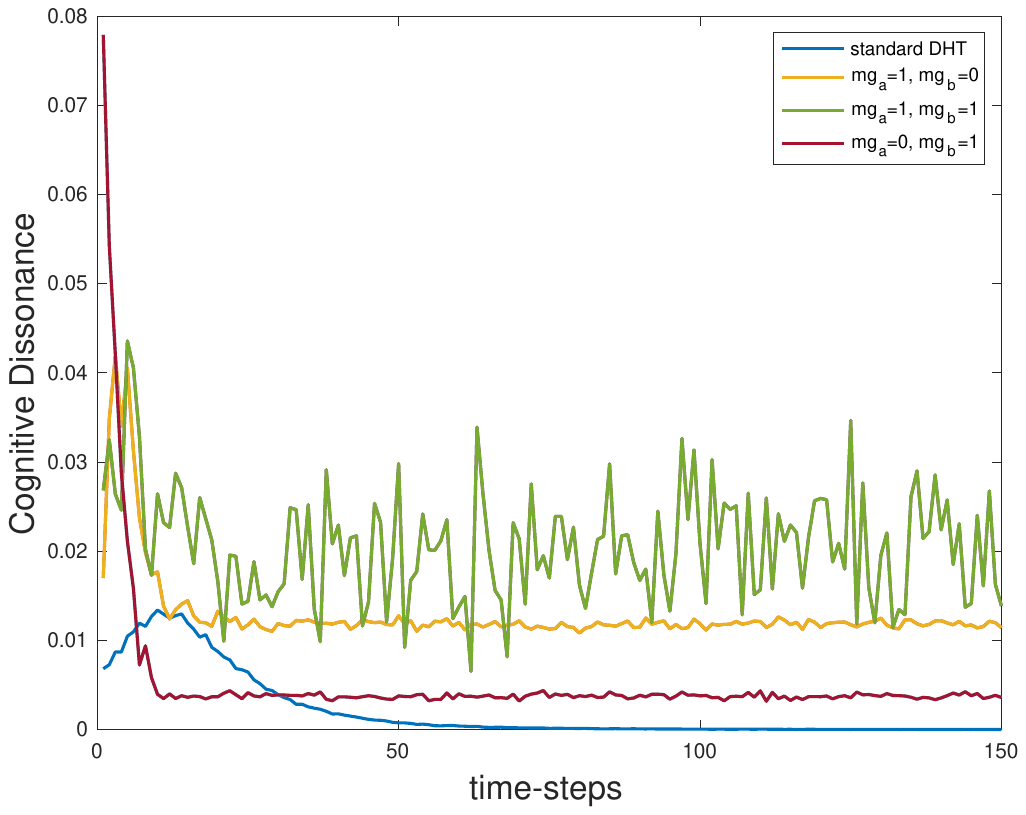}}
        \vspace*{8pt}
    \caption{CD under `mega-nodes' in a series of BA networks (left) $k=10$, (right) $k=100$  with $N=100$ and $M=4$}
    \label{fig:10}
\end{figure}

\begin{figure}[h!]
    \centering
      \centerline{
      \includegraphics[width=7cm]{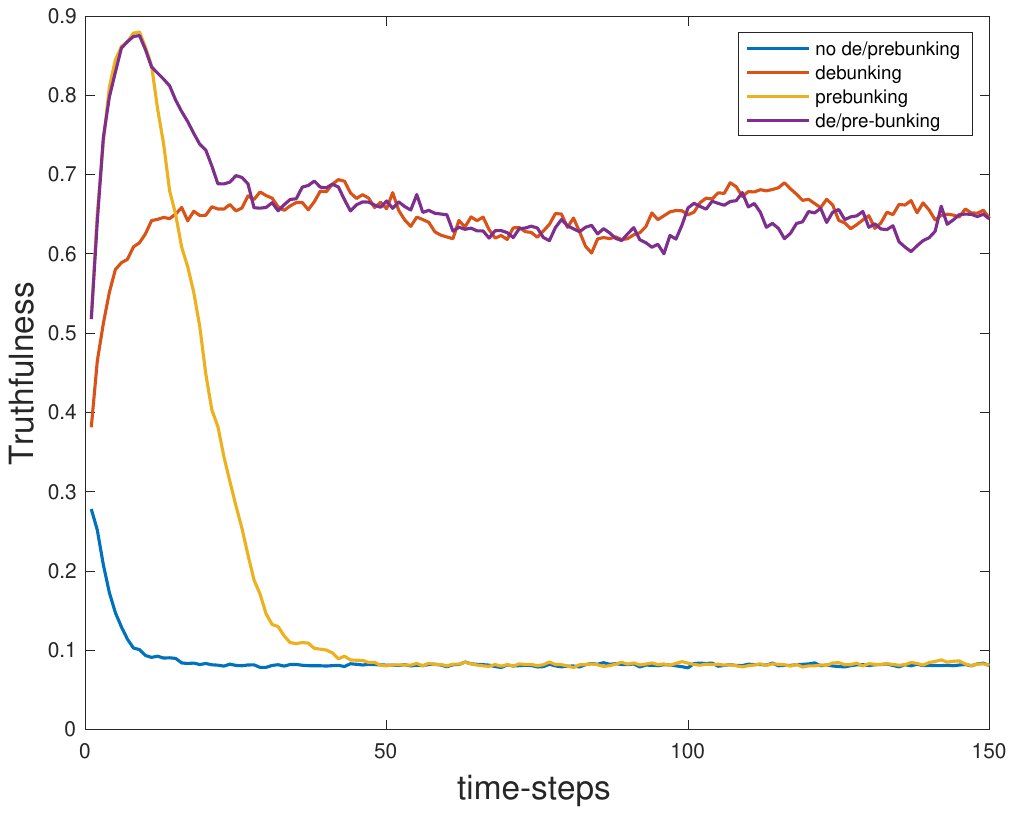}
      \includegraphics[width=7cm]{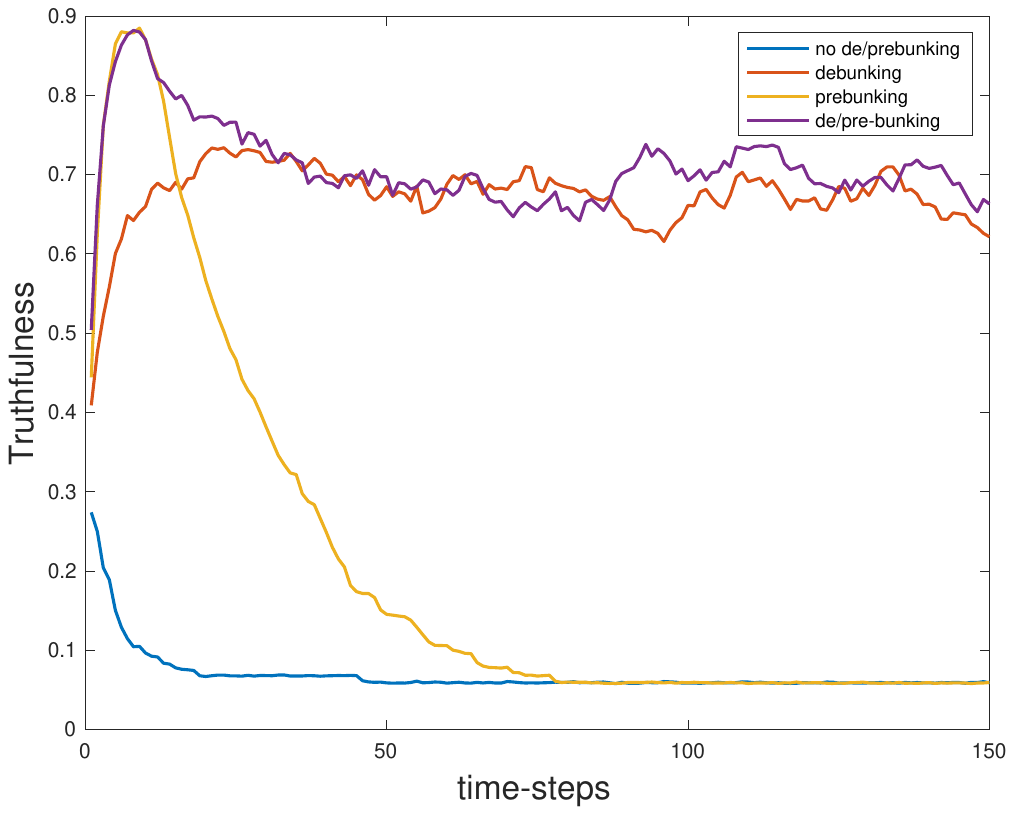}}
       \vspace*{8pt}
    \caption{Effect of Prebunking (centralized) on the temporal evolution of truthfulness with $N=100$ and $M=4$ in a series of BA networks (left) $k=10$, (right) $k=100$}
    \label{fig:11}
\end{figure}

\end{document}